\documentclass{jtacs}

\usepackage{arydshln}
\usepackage{layout}
\usepackage{amsmath}
\usepackage{textcomp}
\usepackage{adjustbox}
\usepackage{graphicx}
\usepackage{placeins}

 \numberwithin{equation}{section}

\begin{document}

%-------------------------------------------------------------------------
% editorial commands: to be inserted by the editorial office
%
%\firstpage{1} \volume{228} \Copyrightyear{2004} \DOI{003-0001}
%\seriesextraline{This is the Concrete Title of this Book\br H.E. R and S.T.C. W, Eds.}
%\firstpage{1}

\title{An algorithm for multipication of Kaluza numbers}
\maketitle
%----------Author 1
\xauthor{A.~Cariow}

\xaffiliation{West Pomeranian University of Technology, Szczecin, \\
Faculty of Computer Science and Information Technology, \\
\.{Z}o\l{}nierska 49, 71-210 Szczecin, Poland}

\xemail{acariow@wi.zut.edu.pl}

%\thanks{This work was completed with the support of our
%----------Author 2
\xauthor{G.~Cariowa}
\xaffiliation{West Pomeranian University of Technology, Szczecin, \\
Faculty of Computer Science and Information Technology, \\
\.{Z}o\l{}nierska 49, 71-210 Szczecin, Poland}
\xemail{gcariowa@wi.zut.edu.pl}

\xauthor{R.~\L{}entek}
\xaffiliation{West Pomeranian University of Technology, Szczecin, \\
Faculty of Computer Science and Information Technology, \\
\.{Z}o\l{}nierska 49, 71-210 Szczecin, Poland}
\xemail{rlentek@wi.zut.edu.pl}
%\subjclass{Primary 65F30; Secondary 15A66}

%\end{tabular}
%\date{September 1, 2014}

\begin{abstract}
This paper presents the derivation of a new algorithm for multiplying of two Kaluza numbers.
Performing this operation directly requires 1024 real multiplications and 992 real additions.  The proposed algorithm can compute the same result with only 512 real multiplications and 576 real additions. The derivation of our algorithm is based on utilizing the fact that multiplication of two Kaluza numbers can be expressed as a matrix–vector product. The matrix multiplicand that participates in the product calculating has unique structural properties. Namely exploitation of these specific properties leads to significant reducing of the complexity of Kaluza numbers multiplication.

\end{abstract}

%%% ----------------------------------------------------------------------

%%% ----------------------------------------------------------------------
%\tableofcontents

%% ###################################################################

\section{Introduction}

Today hypercomplex numbers \cite{1} are used in various fields of data processing including digital signal and image processing, machine graphics, telecommunications and cryptography \cite{2,3,4,5,6,7,8,9,10}. The most popular are quaternions, octonions and sedenions \cite{1}. Perhaps the less popular are the Pauli, Dirac and Kaluza numbers. This numbers are mostly used in solving different physical problems in electrodynamics, field
theory, etc. Somehow or other, hypercomplex number systems and their applications to data processing are beautiful enough to be worth studying simply for the pleasure of it.

The multiplication of two hypercomplex numbers is often one of the more time consuming operations. The reason for this is, because the addition of $N$-dimensional hypercomplex
numbers requires $N$ real additions, while the multiplication of these numbers already
requires $N(N-1)$ real additions and $N^2$ real multiplication. It is easy to see that the increasing
of dimensions of hypercomplex number increases the computational complexity of the
multiplication. Therefore, reducing the computational complexity of the multiplication of
hypercomplex numbers is an important scientific and engineering problem.

Efficient algorithms for the multiplication of various hypercomplex numbers already exist \cite{12,13,14,15,16,17,18,19,20,21,22,23,24}. No such algorithms for the multiplication of Kaluza numbers have been proposed. The aim of the present paper is to suggest an efficient algorithm for this purpose.

\section{Preliminary Remarks}

A Kaluza number is defined as follows:

%\begin{equation}
%\mathbf{Y}_{32\times1}=\mathbf{B}_{32}\mathbf{X}_{32\times1}
%\label{e1}
%\end{equation}
\begin{equation}
d=d_{0}+\sum_{n=1}^{31}d_ni_n ,
\label{e1}
\end{equation}
\begin{adjustbox}{margin = 0 10 0 10}
\\
\end{adjustbox}
where $\{d_0\}$  and  $\{d_n\}$, $n=1,\ldots,31$ are real numbers, and
$\{i_n\}$, $n=1,\ldots,31$ are the imaginary units. The results of all possible products of Kaluza numbers imaginary units can be summarized in the following table:

\begin{adjustbox}{margin = 0 20 0 25}
\\
\end{adjustbox}

%@@@@@@@@@@@@@@@@@@@@@@@@@@@@@@@@@@@@@@@@@@@@@@@@@@@@@@@@@@@ FLOAT TABLE-11111111111
\begin{table}[h]
\caption{North-west quadrant of the table for Kaluza numbers imaginary units multiplication}
\begin{adjustbox}{width=0.99\textwidth, margin = 0 0 0 10}
$
\begin{array}{|c||cccccccccccccccc|}
\hline
\times & 1 & e_{1} & e_{2} & e_{3} & e_{4} & e_{5} & e_{6} & e_{7} & e_{8} & e_{9} & e_{10} & e_{11} & e_{12} & e_{13} & e_{14} & e_{15}
\\ \hline\hline
1 & 1 & e_{1} & e_{2} & e_{3} & e_{4} & e_{5} & e_{6} & e_{7} & e_{8} & e_{9} & e_{10} & e_{11} & e_{12} & e_{13} & e_{14} & e_{15}
\\
e_{1} & e_{1} & 1 & e_{6} & e_{7} & e_{8} & e_{9} & e_{2} & e_{3} & e_{4} & e_{5} & e_{16} & e_{17} & e_{18} & e_{19} & e_{20} & e_{21}
\\
e_{2} & e_{2} & -e_{6} & 1 & e_{10} & e_{11} & e_{12} & -e_{1} & -e_{16} & -e_{17} & -e_{18} & e_{3} & e_{4} & e_{5} & e_{22} & e_{23} & e_{24}
\\
e_{3} & e_{3} & -e_{7} & -e_{10} & -1 & e_{13} & e_{14} & e_{16} & e_{1} & -e_{19} & -e_{20} & e_{2} & -e_{22} & -e_{23} & -e_{4} & -e_{5} & e_{25}
\\
e_{4} & e_{4} & -e_{8} & -e_{11} & -e_{13} & -1 & e_{15} & e_{17} & e_{19} & e_{1} & -e_{21} & e_{22} & e_{2} & -e_{24} & e_{3} & -e_{25} & -e_{5}
\\
e_{5} & e_{5} & -e_{9} & -e_{12} & -e_{14} & -e_{15} & -1 & e_{18} & e_{20} & e_{21} & e_{1} & e_{23} & e_{24} & e_{2} & e_{25} & e_{3} & e_{4}
\\
e_{6} & e_{6} & -e_{2} & e_{1} & e_{16} & e_{17} & e_{18} & -1 & -e_{10} & -e_{11} & -e_{12} & e_{7} & e_{8} & e_{9} & e_{26} & e_{27} & e_{28}
\\
e_{7} & e_{7} & -e_{3} & -e_{16} & -e_{1} & e_{19} & e_{20} & e_{10} & 1 & -e_{13} & -e_{14} & e_{6} & -e_{26} & -e_{27} & -e_{8} & -e_{9} & e_{29}
\\
e_{8} & e_{8} & -e_{4} & -e_{17} & -e_{19} & -e_{1} & e_{21} & e_{11} & e_{13} & 1 & -e_{15} & e_{26} & e_{6} & -e_{28} & e_{7} & -e_{29} & -e_{9}
\\
e_{9} & e_{9} & -e_{5} & -e_{18} & -e_{20} & -e_{21} & -e_{1} & e_{12} & e_{14} & e_{15} & 1 & e_{27} & e_{28} & e_{6} & e_{29} & e_{7} & e_{8}
\\
e_{10} & e_{10} & e_{16} & -e_{3} & -e_{2} & e_{22} & e_{23} & -e_{7} & -e_{6} & e_{26} & e_{27} & 1 & -e_{13} & -e_{14} & -e_{11} & -e_{12} & e_{30}
\\
e_{11} & e_{11} & e_{17} & -e_{4} & -e_{22} & -e_{2} & e_{24} & -e_{8} & -e_{26} & -e_{6} & e_{28} & e_{13} & 1 & -e_{15} & e_{10} & -e_{30} & -e_{12}
\\
e_{12} & e_{12} & e_{18} & -e_{5} & -e_{23} & -e_{24} & -e_{2} & -e_{9} & -e_{27} & -e_{28} & -e_{6} & e_{14} & e_{15} & 1 & e_{30} & e_{10} & e_{11}
\\
e_{13} & e_{13} & e_{19} & e_{22} & e_{4} & -e_{3} & e_{25} & e_{26} & e_{8} & -e_{7} & e_{29} & e_{11} & -e_{10} & e_{30} & -1 & e_{15} & -e_{14}
\\
e_{14} & e_{14} & e_{20} & e_{23} & e_{5} & -e_{25} & -e_{3} & e_{27} & e_{9} & -e_{29} & -e_{7} & e_{12} & -e_{30} & -e_{10} & -e_{15} & -1 & e_{13}
\\
e_{15} & e_{15} & e_{21} & e_{24} & e_{25} & e_{5} & -e_{4} & e_{28} & e_{29} & e_{9} & -e_{8} & e_{30} & e_{12} & -e_{11} & e_{14} & -e_{13} & -1
\\ \hline
\end{array}
$
\end{adjustbox}
\label{tab1}
\end{table}%ZROBIONE (KALUZA)
%@@@@@@@@@@@@@@@@@@@@@@@@@@@@@@@@@@@@@@@@@@@@@@@@@@@@@@@@@@@
%@@@@@@@@@@@@@@@@@@@@@@@@@@@@@@@@@@@@@@@@@@@@@@@@@@@@@@@@@@@ FLOAT TABLE-2222222222222 PIerwsza zrobiona!!!!!!!!!!!!!!
\begin{table}[h]
\caption{North-east quadrant of the table for Kaluza numbers imaginary units multiplication}
\begin{adjustbox}{width=0.99\textwidth}
$
\begin{array}{|c||cccccccccccccccc|}
\hline
\times & e_{16} & e_{17} & e_{18} & e_{19} & e_{20} & e_{21} & e_{22} & e_{23} & e_{24} & e_{25} & e_{26} & e_{27} & e_{28} & e_{29} & e_{30} & e_{31}
\\ \hline\hline
1 & e_{16} & e_{17} & e_{18} & e_{19} & e_{20} & e_{21} & e_{22} & e_{23} & e_{24} & e_{25} & e_{26} & e_{27} & e_{28} & e_{29} & e_{30} & e_{31}
\\
e_{1} & e_{10} & e_{11} & e_{12} & e_{13} & e_{14} & e_{15} & e_{26} & e_{27} & e_{28} & e_{29} & e_{22} & e_{23} & e_{24} & e_{25} & e_{31} & e_{30}
\\
e_{2} & -e_{7} & -e_{8} & -e_{9} & -e_{26} & -e_{27} & -e_{28} & -e_{13} & e_{14} & e_{15} & e_{30} & -e_{19} & -e_{20} & -e_{21} & -e_{31} & e_{25} & -e_{29}
\\
e_{3} & -e_{6} & e_{26} & e_{27} & e_{8} & e_{9} & -e_{29} & e_{11} & e_{12} & -e_{30} & -e_{15} & -e_{17} & -e_{18} & e_{31} & e_{21} & e_{24} & -e_{28}
\\
e_{4} & -e_{26} & -e_{6} & e_{28} & -e_{7} & e_{29} & e_{9} & -e_{10} & e_{30} & e_{12} & e_{14} & e_{16} & -e_{31} & -e_{18} & -e_{20} & -e_{23} & e_{27}
\\
e_{5} & -e_{27} & -e_{28} & -e_{6} & -e_{29} & -e_{7} & -e_{8} & -e_{30} & -e_{10} & -e_{11} & -e_{13} & e_{31} & e_{16} & e_{17} & e_{19} & e_{22} & -e_{26}
\\
e_{6} & -e_{3} & -e_{4} & -e_{5} & -e_{22} & -e_{23} & -e_{24} & e_{19} & e_{20} & e_{21} & e_{31} & -e_{13} & -e_{14} & -e_{15} & -e_{30} & e_{29} & -e_{25}
\\
e_{7} & -e_{2} & e_{22} & e_{23} & e_{4} & e_{5} & -e_{25} & e_{17} & e_{18} & -e_{31} & -e_{21} & -e_{11} & -e_{12} & e_{30} & e_{15} & e_{28} & -e_{24}
\\
e_{8} & -e_{22} & -e_{2} & e_{24} & -e_{3} & e_{25} & e_{5} & -e_{16} & e_{31} & e_{18} & e_{20} & e_{10} & -e_{30} & -e_{12} & -e_{14} & -e_{27} & e_{23}
\\
e_{9} & -e_{23} & -e_{24} & -e_{2} & -e_{25} & -e_{3} & -e_{4} & -e_{31} & -e_{16} & -e_{17} & -e_{19} & e_{30} & e_{10} & e_{11} & e_{13} & e_{26} & -e_{22}
\\
e_{10} & e_{1} & -e_{19} & -e_{20} & -e_{17} & -e_{18} & e_{31} & e_{4} & e_{5} & -e_{25} & -e_{24} & e_{8} & e_{9} & -e_{29} & -e_{28} & e_{15} & e_{21}
\\
e_{11} & e_{19} & e_{1} & -e_{21} & e_{16} & -e_{31} & -e_{18} & -e_{3} & e_{25} & e_{5} & e_{23} & -e_{7} & e_{29} & e_{9} & e_{27} & -e_{14} & -e_{20}
\\
e_{12} & e_{20} & e_{21} & e_{1} & e_{31} & e_{16} & e_{17} & -e_{25} & -e_{3} & -e_{4} & -e_{22} & -e_{29} & -e_{7} & -e_{8} & -e_{26} & e_{13} & e_{19}
\\
e_{13} & e_{17} & -e_{16} & e_{31} & -e_{1} & e_{21} & -e_{20} & -e_{2} & e_{24} & -e_{23} & -e_{5} & -e_{6} & e_{28} & -e_{27} & -e_{9} & -e_{12} & -e_{18}
\\
e_{14} & e_{18} & -e_{31} & -e_{16} & -e_{21} & -e_{1} & e_{19} & -e_{24} & -e_{2} & e_{22} & e_{4} & -e_{28} & -e_{6} & e_{26} & e_{8} & e_{11} & e_{17}
\\
e_{15} & e_{31} & e_{18} & -e_{17} & e_{20} & -e_{19} & -e_{1} & e_{23} & -e_{22} & -e_{2} & -e_{3} & e_{27} & -e_{26} & -e_{6} & -e_{7} & -e_{10} & -e_{16}
\\ \hline
\end{array}
$
\end{adjustbox}
\label{tab2}
\end{table}
%@@@@@@@@@@@@@@@@@@@@@@@@@@@@@@@@@@@@@@@@@@@@@@@@@@@@@@@@@@@
%@@@@@@@@@@@@@@@@@@@@@@@@@@@@@@@@@@@@@@@@@@@@@@@@@@@@@@@@@@@ FLOAT TABLE-33333333333 TABELA2 SKONCZONA!!!!!!!!!!!!
\begin{table}[h]
\caption{South-west quadrant of the table for Kaluza numbers imaginary units multiplication}
\begin{adjustbox}{width=0.99\textwidth}
$
\begin{array}{|c||cccccccccccccccc|}
\hline
\times & 1 & e_{1} & e_{2} & e_{3} & e_{4} & e_{5} & e_{6} & e_{7} & e_{8} & e_{9} & e_{10} & e_{11} & e_{12} & e_{13} & e_{14} & e_{15}
\\ \hline\hline
e_{16} & e_{16} & e_{10} & -e_{7} & -e_{6} & e_{26} & e_{27} & -e_{3} & -e_{2} & e_{22} & e_{23} & e_{1} & -e_{19} & -e_{20} & -e_{17} & -e_{18} & e_{31}
\\
e_{17} & e_{17} & e_{11} & -e_{8} & -e_{26} & -e_{6} & e_{28} & -e_{4} & -e_{22} & -e_{2} & e_{24} & e_{19} & e_{1} & -e_{21} & e_{16} & -e_{31} & -e_{18}
\\
e_{18} & e_{18} & e_{12} & -e_{9} & -e_{27} & -e_{28} & -e_{6} & -e_{5} & -e_{23} & -e_{24} & -e_{2} & e_{20} & e_{21} & e_{1} & e_{31} & e_{16} & e_{17}
\\
e_{19} & e_{19} & e_{13} & e_{26} & e_{8} & -e_{7} & e_{29} & e_{22} & e_{4} & -e_{3} & e_{25} & e_{17} & -e_{16} & e_{31} & -e_{1} & e_{21} & -e_{20}
\\
e_{20} & e_{20} & e_{14} & e_{27} & e_{9} & -e_{29} & -e_{7} & e_{23} & e_{5} & -e_{25} & -e_{3} & e_{18} & -e_{31} & -e_{16} & -e_{21} & -e_{1} & e_{19}
\\
e_{21} & e_{21} & e_{15} & e_{28} & e_{29} & e_{9} & -e_{8} & e_{24} & e_{25} & e_{5} & -e_{4} & e_{31} & e_{18} & -e_{17} & e_{20} & -e_{19} & -e_{1}
\\
e_{22} & e_{22} & -e_{26} & e_{13} & e_{11} & -e_{10} & e_{30} & -e_{19} & -e_{17} & e_{16} & -e_{31} & e_{4} & -e_{3} & e_{25} & -e_{2} & e_{24} & -e_{23}
\\
e_{23} & e_{23} & -e_{27} & e_{14} & e_{12} & -e_{30} & -e_{10} & -e_{20} & -e_{18} & e_{31} & e_{16} & e_{5} & -e_{25} & -e_{3} & -e_{24} & -e_{2} & e_{22}
\\
e_{24} & e_{24} & -e_{28} & e_{15} & e_{30} & e_{12} & -e_{11} & -e_{21} & -e_{31} & -e_{18} & e_{17} & e_{25} & e_{5} & -e_{4} & e_{23} & -e_{22} & -e_{2}
\\
e_{25} & e_{25} & -e_{29} & -e_{30} & -e_{15} & e_{14} & -e_{13} & e_{31} & e_{21} & -e_{20} & e_{19} & e_{24} & -e_{23} & e_{22} & -e_{5} & e_{4} & -e_{3}
\\
e_{26} & e_{26} & -e_{22} & e_{19} & e_{17} & -e_{16} & e_{31} & -e_{13} & -e_{11} & e_{10} & -e_{30} & e_{8} & -e_{7} & e_{29} & -e_{6} & e_{28} & -e_{27}
\\
e_{27} & e_{27} & -e_{23} & e_{20} & e_{18} & -e_{31} & -e_{16} & -e_{14} & -e_{12} & e_{30} & e_{10} & e_{9} & -e_{29} & -e_{7} & -e_{28} & -e_{6} & e_{26}
\\
e_{28} & e_{28} & -e_{24} & e_{21} & e_{31} & e_{18} & -e_{17} & -e_{15} & -e_{30} & -e_{12} & e_{11} & e_{29} & e_{9} & -e_{8} & e_{27} & -e_{26} & -e_{6}
\\
e_{29} & e_{29} & -e_{25} & -e_{31} & -e_{21} & e_{20} & -e_{19} & e_{30} & e_{15} & -e_{14} & e_{13} & e_{28} & -e_{27} & e_{26} & -e_{9} & e_{8} & -e_{7}
\\
e_{30} & e_{30} & e_{31} & -e_{25} & -e_{24} & e_{23} & -e_{22} & -e_{29} & -e_{28} & e_{27} & -e_{26} & e_{15} & -e_{14} & e_{13} & -e_{12} & e_{11} & -e_{10}
\\
e_{31} & e_{31} & e_{30} & -e_{29} & -e_{28} & e_{27} & -e_{26} & -e_{25} & -e_{24} & e_{23} & -e_{22} & e_{21} & -e_{20} & e_{19} & -e_{18} & e_{17} & -e_{16}
\\ \hline
\end{array}
$
\end{adjustbox}
\label{tab3}
\end{table}
%@@@@@@@@@@@@@@@@@@@@@@@@@@@@@@@@@@@@@@@@@@@@@@@@@@@@@@@@@@@
%@@@@@@@@@@@@@@@@@@@@@@@@@@@@@@@@@@@@@@@@@@@@@@@@@@@@@@@@@@@ FLOAT TABLE-4444444444444 Tabela 3 skonczona!!!!!!!!
\begin{table}[h]%przestawic jakos
\caption{South-east quadrant of the table for Kaluza numbers imaginary units multiplication }
\begin{adjustbox}{width=0.99\textwidth}
$
\begin{array}{|c||cccccccccccccccc|}
\hline
\times & e_{16} & e_{17} & e_{18} & e_{19} & e_{20} & e_{21} & e_{22} & e_{23} & e_{24} & e_{25} & e_{26} & e_{27} & e_{28} & e_{29} & e_{30} & e_{31}
\\ \hline\hline
e_{16} & 1 & -e_{13} & -e_{14} & -e_{11} & -e_{12} & e_{30} & e_{8} & e_{9} & -e_{29} & -e_{28} & e_{4} & e_{5} & -e_{25} & -e_{24} & e_{21} & e_{15}
\\
e_{17} & e_{13} & 1 & -e_{15} & e_{10} & -e_{30} & -e_{12} & -e_{7} & e_{29} & e_{9} & e_{27} & -e_{3} & e_{25} & e_{5} & e_{23} & -e_{20} & -e_{14}
\\
e_{18} & e_{14} & e_{15} & 1 & e_{30} & e_{10} & e_{11} & -e_{29} & -e_{7} & -e_{8} & -e_{26} & -e_{25} & -e_{3} & -e_{4} & -e_{22} & e_{19} & e_{13}
\\
e_{19} & e_{11} & -e_{10} & e_{30} & -1 & e_{15} & -e_{14} & -e_{6} & e_{28} & -e_{27} & -e_{9} & -e_{2} & e_{24} & -e_{23} & -e_{5} & -e_{18} & -e_{12}
\\
e_{20} & e_{12} & -e_{30} & -e_{10} & -e_{15} & -1 & e_{13} & -e_{28} & -e_{6} & e_{26} & e_{8} & -e_{24} & -e_{2} & e_{22} & e_{4} & e_{17} & e_{11}
\\
e_{21} & e_{30} & e_{12} & -e_{11} & e_{14} & -e_{13} & -1 & e_{27} & -e_{26} & -e_{6} & -e_{7} & e_{23} & -e_{22} & -e_{2} & -e_{3} & -e_{16} & -e_{10}
\\
e_{22} & -e_{8} & e_{7} & -e_{29} & e_{6} & -e_{28} & e_{27} & -1 & e_{15} & -e_{14} & -e_{12} & e_{1} & -e_{21} & e_{20} & e_{18} & -e_{5} & e_{9}
\\
e_{23} & -e_{9} & e_{29} & e_{7} & e_{28} & e_{6} & -e_{26} & -e_{15} & -1 & e_{13} & e_{11} & e_{21} & e_{1} & -e_{19} & -e_{17} & e_{4} & -e_{8}
\\
e_{24} & -e_{29} & -e_{9} & e_{8} & -e_{27} & e_{26} & e_{6} & e_{14} & -e_{13} & -1 & -e_{10} & -e_{20} & e_{19} & e_{1} & e_{16} & -e_{3} & e_{7}
\\
e_{25} & -e_{28} & e_{27} & -e_{26} & e_{9} & -e_{8} & e_{7} & e_{12} & -e_{11} & e_{10} & 1 & -e_{18} & e_{17} & -e_{16} & -e_{1} & -e_{2} & e_{6}
\\
e_{26} & -e_{4} & e_{3} & -e_{25} & e_{2} & -e_{24} & e_{23} & -e_{1} & e_{21} & -e_{20} & -e_{18} & 1 & -e_{15} & e_{14} & e_{12} & -e_{9} & e_{5}
\\
e_{27} & -e_{5} & e_{25} & e_{3} & e_{24} & e_{2} & -e_{22} & -e_{21} & -e_{1} & e_{19} & e_{17} & e_{15} & 1 & -e_{13} & -e_{11} & e_{8} & -e_{4}
\\
e_{28} & -e_{25} & -e_{5} & e_{4} & -e_{23} & e_{22} & e_{2} & e_{20} & -e_{19} & -e_{1} & -e_{16} & -e_{14} & e_{13} & 1 & e_{10} & -e_{7} & e_{3}
\\
e_{29} & -e_{24} & e_{23} & -e_{22} & e_{5} & -e_{4} & e_{3} & e_{18} & -e_{17} & e_{16} & e_{1} & -e_{12} & e_{11} & -e_{10} & -1 & -e_{6} & e_{2}
\\
e_{30} & e_{21} & -e_{20} & e_{19} & -e_{18} & e_{17} & -e_{16} & e_{5} & -e_{4} & e_{3} & e_{2} & e_{9} & -e_{8} & e_{7} & e_{6} & -1 & -e_{1}
\\
e_{31} & e_{15} & -e_{14} & e_{13} & -e_{12} & e_{11} & -e_{10} & e_{9} & -e_{8} & e_{7} & e_{6} & e_{5} & -e_{4} & e_{3} & e_{2} & -e_{1} & -1
\\ \hline
\end{array}
$
\end{adjustbox}
\label{tab4}
\end{table}
%@@@@@@@@@@@@@@@@@@@@@@@@@Tabela 4 skonczona@@@@@@@@@@@@@@@@@@@
\FloatBarrier
Suppose we want to compute the product of two Kaluza numbers.

\begin{equation}
d=d^{(1)}d^{(2)}=d_0+\sum_{n=1}^{31}d_ni_n ,
\label{2}
\end{equation}
where
\[
d^{(1)}=a_0+\sum_{n=1}^{31}a_ni_n , d^{(2)}=b_0+\sum_{n=1}^{31}b_ni_n .
\]
%przestawia sie!!!!!!!!!!!!!!!!!!!!!!!!!!!!!!!!!!!!!!!! begin{equation} numeruje równania z boku

The operation of multiplication of Kaluza numbers can be represented more compactly in
the form of vector-matrix product:
\begin{equation}
\mathbf{Y}_{32\times1}=\mathbf{B}_{32}\mathbf{X}_{32\times1} ,
\label{2}
\end{equation}
where
\[
\mathbf{X}_{32\times1}=[a_0,a_1,\ldots,a_{31}]^{\mathrm{T}},
\]
\[
\mathbf{Y}_{32\times1}=[d_0,d_1,\ldots,d_{15}]^{\mathrm{T}},
\]

\[
\mathbf{B}_{32}=\left[
\begin{array}{ll}
\mathbf{B}_{16}^{(0,0)} & \mathbf{B}_{16}^{(1,0)} \\
\mathbf{B}_{16}^{(0,1)} & \mathbf{B}_{16}^{(1,1)}
\end{array}\right],
\]

%%%%%%%%%%%%%%%%%%%%%%%%%%%%%%%%%%%%%%%%%%%%%%%TABELA
$\mathbf{B}_{16}^{(0,0)}=$\\*
\begin{adjustbox}{width=\textwidth, margin= 3 15 0 5}
$
=\left[
\begin{array}{cccccccccccccccc}
\begin{matrix}
b_{0} & b_{1} & b_{2} & -b_{3} & -b_{4} & -b_{5} & -b_{6} & b_{7} \\
b_{1} & b_{0} & -b_{6} & b_{7} & b_{8} & b_{9} & b_{2} & -b_{3} \\
b_{2} & b_{6} & b_{0} & b_{10} & b_{11} & b_{12} & -b_{1} & -b_{16} \\
b_{3} & b_{7} & b_{10} & b_{0} & b_{13} & b_{14} & -b_{16} & -b_{1} \\
b_{4} & b_{8} & b_{11} & -b_{13} & b_{0} & b_{15} & -b_{17} & b_{19} \\
b_{5} & b_{9} & b_{12} & -b_{14} & -b_{15} & b_{0} & -b_{18} & b_{20} \\
b_{6} & b_{2} & -b_{1} & -b_{16} & -b_{17} & -b_{18} & b_{0} & b_{10} \\
b_{7} & b_{3} & -b_{16} & -b_{1} & -b_{19} & -b_{20} & b_{10} & b_{0} \\
b_{8} & b_{4} & -b_{17} & b_{19} & -b_{1} & -b_{21} & b_{11} & -b_{13} \\
b_{9} & b_{5} & -b_{18} & b_{20} & b_{21} & -b_{1} & b_{12} & -b_{14} \\
b_{10} & b_{16} & b_{3} & -b_{2} & -b_{22} & -b_{23} & -b_{7} & b_{6} \\
b_{11} & b_{17} & b_{4} & b_{22} & -b_{2} & -b_{24} & -b_{8} & -b_{26} \\
b_{12} & b_{18} & b_{5} & b_{23} & b_{24} & -b_{2} & -b_{9} & -b_{27} \\
b_{13} & b_{19} & b_{22} & b_{4} & -b_{3} & -b_{25} & -b_{26} & -b_{8} \\
b_{14} & b_{20} & b_{23} & b_{5} & b_{25} & -b_{3} & -b_{27} & -b_{9} \\
b_{15} & b_{21} & b_{24} & -b_{25} & b_{5} & -b_{4} & -b_{28} & b_{29} \\
\end{matrix} &
\begin{matrix}
b_{8} & b_{9} & b_{10} & b_{11} & b_{12} & -b_{13} & -b_{14} & -b_{15} \\
-b_{4} & -b_{5} & b_{16} & b_{17} & b_{18} & -b_{19} & -b_{20} & -b_{21} \\
-b_{17} & -b_{18} & -b_{3} & -b_{4} & -b_{5} & -b_{22} & -b_{23} & -b_{24} \\
-b_{19} & -b_{20} & -b_{2} & -b_{22} & -b_{23} & -b_{4} & -b_{5} & -b_{25} \\
-b_{1} & -b_{21} & b_{22} & -b_{2} & -b_{24} & b_{3} & b_{25} & -b_{5} \\
b_{21} & -b_{1} & b_{23} & b_{24} & -b_{2} & -b_{25} & b_{3} & b_{4} \\
b_{11} & b_{12} & -b_{7} & -b_{8} & -b_{9} & -b_{26} & -b_{27} & -b_{28} \\
b_{13} & b_{14} & -b_{6} & -b_{26} & -b_{27} & -b_{8} & -b_{9} & -b_{29} \\
b_{0} & b_{15} & b_{26} & -b_{6} & -b_{28} & b_{7} & b_{29} & -b_{9} \\
-b_{15} & b_{0} & b_{27} & b_{28} & -b_{6} & -b_{29} & b_{7} & b_{8} \\
b_{26} & b_{27} & b_{0} & b_{13} & b_{14} & -b_{11} & -b_{12} & -b_{30} \\
b_{6} & b_{28} & -b_{13} & b_{0} & b_{15} & b_{10} & b_{30} & -b_{12} \\
-b_{28} & b_{6} & -b_{14} & -b_{15} & b_{0} & -b_{30} & b_{10} & b_{11} \\
b_{7} & b_{29} & -b_{11} & b_{10} & b_{30} & b_{0} & b_{15} & -b_{14} \\
-b_{29} & b_{7} & -b_{12} & -b_{30} & b_{10} & -b_{15} & b_{0} & b_{13} \\
-b_{9} & b_{8} & b_{30} & -b_{12} & b_{11} & b_{14} & -b_{13} & b_{0}\\
\end{matrix}
\end{array}
\right],
$
\end{adjustbox}
%%%NW gotowa
$\mathbf{B}_{16}^{(0,1)}=$\\*
\begin{adjustbox}{width=\textwidth, margin= 3 15 0 5}
$
=\left[
\begin{array}{cc}
\begin{matrix}
b_{16} & b_{10} & -b_{7} & b_{6} & b_{26} & b_{27} & b_{3} & -b_{2} \\
b_{17} & b_{11} & -b_{8} & -b_{26} & b_{6} & b_{28} & b_{4} & b_{22} \\
b_{18} & b_{12} & -b_{9} & -b_{27} & -b_{28} & b_{6} & b_{5} & b_{23} \\
b_{19} & b_{13} & -b_{26} & -b_{8} & b_{7} & b_{29} & b_{22} & b_{4} \\
b_{20} & b_{14} & -b_{27} & -b_{9} & -b_{29} & b_{7} & b_{23} & b_{5} \\
b_{21} & b_{15} & -b_{28} & b_{29} & -b_{9} & b_{8} & b_{24} & -b_{25} \\
b_{22} & b_{26} & b_{13} & -b_{11} & b_{10} & b_{30} & -b_{19} & b_{17} \\
b_{23} & b_{27} & b_{14} & -b_{12} & -b_{30} & b_{10} & -b_{20} & b_{18} \\
b_{24} & b_{28} & b_{15} & b_{30} & -b_{12} & b_{11} & -b_{21} & -b_{31} \\
b_{25} & b_{29} & b_{30} & b_{15} & -b_{14} & b_{13} & -b_{31} & -b_{21} \\
b_{26} & b_{22} & -b_{19} & b_{17} & -b_{16} & -b_{31} & b_{13} & -b_{11} \\
b_{27} & b_{23} & -b_{20} & b_{18} & b_{31} & -b_{16} & b_{14} & -b_{12} \\
b_{28} & b_{24} & -b_{21} & -b_{31} & b_{18} & -b_{17} & b_{15} & b_{30} \\
b_{29} & b_{25} & -b_{31} & -b_{21} & b_{20} & -b_{19} & b_{30} & b_{15} \\
 b_{30} & b_{31} & b_{25} & -b_{24} & b_{23} & -b_{22} & -b_{29} & b_{28} \\
b_{31} & b_{30} & -b_{29} & b_{28} & -b_{27} & b_{26} & b_{25} & -b_{24} \\
\end{matrix} &
\begin{matrix}
-b_{22} & -b_{23} & b_{1} & b_{19} & b_{20} & -b_{17} & -b_{18} & -b_{31} \\
-b_{2} & -b_{24} & -b_{19} & b_{1} & b_{21} & b_{16} & b_{31} & -b_{18} \\
b_{24} & -b_{2} & -b_{20} & -b_{21} & b_{1} & -b_{31} & b_{16} & b_{17} \\
-b_{3} & -b_{25} & -b_{17} & b_{16} & b_{31} & b_{1} & b_{21} & -b_{20} \\
b_{25} & -b_{3} & -b_{18} & -b_{31} & b_{16} & -b_{21} & b_{1} & b_{19} \\
b_{5} & -b_{4} & b_{31} & -b_{18} & b_{17} & b_{20} & -b_{19} & b_{1} \\
-b_{16} & -b_{31} & b_{4} & -b_{3} & -b_{25} & b_{2} & b_{24} & -b_{23} \\
b_{31} & -b_{16} & b_{5} & b_{25} & -b_{3} & -b_{24} & b_{2} & b_{22} \\
b_{18} & -b_{17} & -b_{25} & b_{5} & -b_{4} & b_{23} & -b_{22} & b_{2} \\
b_{20} & -b_{19} & -b_{24} & b_{23} & -b_{22} & b_{5} & -b_{4} & b_{3} \\
b_{10} & b_{30} & b_{8} & -b_{7} & -b_{29} & b_{6} & b_{28} & -b_{27} \\
-b_{30} & b_{10} & b_{9} & b_{29} & -b_{7} & -b_{28} & b_{6} & b_{26} \\
-b_{12} & b_{11} & -b_{29} & b_{9} & -b_{8} & b_{27} & -b_{26} & b_{6} \\
-b_{14} & b_{13} & -b_{28} & b_{27} & -b_{26} & b_{9} & -b_{8} & b_{7} \\
-b_{27} & b_{26} & b_{15} & -b_{14} & b_{13} & b_{12} & -b_{11} & b_{10} \\
b_{23} & -b_{22} & b_{21} & -b_{20} & b_{19} & b_{18} & -b_{17} & b_{16} \\
\end{matrix}
\end{array}
\right],
$
\end{adjustbox}
\\
$\mathbf{B}_{16}^{(1,0)}=$\\*
\begin{adjustbox}{width=\textwidth, margin= 3 15 0 5}
$
=\left[
\begin{array}{cc}
\begin{matrix}
b_{16} & b_{17} & b_{18} & -b_{19} & -b_{20} & -b_{21} & -b_{22} & -b_{23} \\
b_{10} & b_{11} & b_{12} & -b_{13} & -b_{14} & -b_{15} & b_{26} & b_{27} \\
-b_{7} & -b_{8} & -b_{9} & -b_{26} & -b_{27} & -b_{28} & -b_{13} & -b_{14} \\
-b_{6} & -b_{26} & -b_{27} & -b_{8} & -b_{9} & -b_{29} & -b_{11} & -b_{12} \\
b_{26} & -b_{6} & -b_{28} & b_{7} & b_{29} & -b_{9} & b_{10} & b_{30} \\
b_{27} & b_{28} & -b_{6} & -b_{29} & b_{7} & b_{8} & -b_{30} & b_{10} \\
-b_{3} & -b_{4} & -b_{5} & -b_{22} & -b_{23} & -b_{24} & b_{19} & b_{20} \\
-b_{2} & -b_{22} & -b_{23} & -b_{4} & -b_{5} & -b_{25} & b_{17} & b_{18} \\
 b_{22} & -b_{2} & -b_{24} & b_{3} & b_{25} & -b_{5} & -b_{16} & -b_{31} \\
b_{23} & b_{24} & -b_{2} & -b_{25} & b_{3} & b_{4} & b_{31} & -b_{16} \\
b_{1} & b_{19} & b_{20} & -b_{17} & -b_{18} & -b_{31} & -b_{4} & -b_{5} \\
-b_{19} & b_{1} & b_{21} & b_{16} & b_{31} & -b_{18} & b_{3} & b_{25} \\
-b_{20} & -b_{21} & b_{1} & -b_{31} & b_{16} & b_{17} & -b_{25} & b_{3} \\
-b_{17} & b_{16} & b_{31} & b_{1} & b_{21} & -b_{20} & b_{2} & b_{24} \\
-b_{18} & -b_{31} & b_{16} & -b_{21} & b_{1} & b_{19} & -b_{24} & b_{2} \\
b_{31} & -b_{18} & b_{17} & b_{20} & -b_{19} & b_{1} & b_{23} & -b_{22} \\
\end{matrix} &
\begin{matrix}
-b_{24} & b_{25} & b_{26} & b_{27} & b_{28} & -b_{29} & -b_{30} & -b_{31} \\
b_{28} & -b_{29} & -b_{22} & -b_{23} & -b_{24} & b_{25} & -b_{31} & -b_{30} \\
-b_{15} & -b_{30} & b_{19} & b_{20} & b_{21} & b_{31} & b_{25} & b_{29} \\
-b_{30} & -b_{15} & b_{17} & b_{18} & b_{31} & b_{21} & b_{24} & b_{28} \\
-b_{12} & b_{14} & -b_{16} & -b_{31} & b_{18} & -b_{20} & -b_{23} & -b_{27} \\
b_{11} & -b_{13} & b_{31} & -b_{16} & -b_{17} & b_{19} & b_{22} & b_{26} \\
b_{21} & b_{31} & -b_{13} & -b_{14} & -b_{15} & -b_{30} & b_{29} & b_{25} \\
b_{31} & b_{21} & -b_{11} & -b_{12} & -b_{30} & -b_{15} & b_{28} & b_{24} \\
b_{18} & -b_{20} & b_{10} & b_{30} & -b_{12} & b_{14} & -b_{27} & -b_{23} \\
-b_{17} & b_{19} & -b_{30} & b_{10} & b_{11} & -b_{13} & b_{26} & b_{22} \\
-b_{25} & b_{24} & b_{8} & b_{9} & b_{29} & -b_{28} & -b_{15} & -b_{21} \\
-b_{5} & -b_{23} & -b_{7} & -b_{29} & b_{9} & b_{27} & b_{14} & b_{20} \\
b_{4} & b_{22} & b_{29} & -b_{7} & -b_{8} & -b_{26} & -b_{13} & -b_{19} \\
-b_{23} & -b_{5} & -b_{6} & -b_{28} & b_{27} & b_{9} & b_{12} & b_{18} \\
b_{22} & b_{4} & b_{28} & -b_{6} & -b_{26} & -b_{8} & -b_{11} & -b_{17} \\
b_{2} & -b_{3} & -b_{27} & b_{26} & -b_{6} & b_{7} & b_{10} & b_{16}\\
\end{matrix}
\end{array}
\right],
$
\end{adjustbox}
\\
$\mathbf{B}_{16}^{(1,1)}=$\\*
\begin{adjustbox}{width=\textwidth, margin= 3 15 0 5}
$
=\left[
\begin{array}{cc}
\begin{matrix}
b_{0} & b_{13} & b_{14} & -b_{11} & -b_{12} & -b_{30} & b_{8} & b_{9} \\
-b_{13} & b_{0} & b_{15} & b_{10} & b_{30} & -b_{12} & -b_{7} & -b_{29}\\
-b_{14} & -b_{15} & b_{0} & -b_{30} & b_{10} & b_{11} & b_{29} & -b_{7} \\
-b_{11} & b_{10} & b_{30} & b_{0} & b_{15} & -b_{14} & -b_{6} & -b_{28} \\
-b_{12} & -b_{30} & b_{10} & -b_{15} & b_{0} & b_{13} & b_{28} & -b_{6} \\
b_{30} & -b_{12} & b_{11} & b_{14} & -b_{13} & b_{0} & -b_{27} & b_{26} \\
b_{8} & -b_{7} & -b_{29} & b_{6} & b_{28} & -b_{27} & b_{0} & b_{15} \\
b_{9} & b_{29} & -b_{7} & -b_{28} & b_{6} & b_{26} & -b_{15} & b_{0} \\
-b_{29} & b_{9} & -b_{8} & b_{27} & -b_{26} & b_{6} & b_{14} & -b_{13} \\
-b_{28} & b_{27} & -b_{26} & b_{9} & -b_{8} & b_{7} & b_{12} & -b_{11} \\
b_{4} & -b_{3} & -b_{25} & b_{2} & b_{24} & -b_{23} & -b_{1} & -b_{21} \\
b_{5} & b_{25} & -b_{3} & -b_{24} & b_{2} & b_{22} & b_{21} & -b_{1} \\
-b_{25} & b_{5} & -b_{4} & b_{23} & -b_{22} & b_{2} & -b_{20} & b_{19} \\
-b_{24} & b_{23} & -b_{22} & b_{5} & -b_{4} & b_{3} & -b_{18} & b_{17} \\
b_{21} & -b_{20} & b_{19} & b_{18} & -b_{17} & b_{16} & b_{5} & -b_{4} \\
b_{15} & -b_{14} & b_{13} & b_{12} & -b_{11} & b_{10} & -b_{9} & b_{8} \\
\end{matrix} &
\begin{matrix}
b_{29} & -b_{28} & -b_{4} & -b_{5} & -b_{25} & b_{24} & -b_{21} & -b_{15} \\
b_{9} & b_{27} & b_{3} & b_{25} & -b_{5} & -b_{23} & b_{20} & b_{14} \\
-b_{8} & -b_{26} & -b_{25} & b_{3} & b_{4} & b_{22} & -b_{19} & -b_{13} \\
b_{27} & b_{9} & b_{2} & b_{24} & -b_{23} & -b_{5} & b_{18} & b_{12} \\
-b_{26} & -b_{8} & -b_{24} & b_{2} & b_{22} & b_{4} & -b_{17} & -b_{11} \\
-b_{6} & b_{7} & b_{23} & -b_{22} & b_{2} & -b_{3} & b_{16} & b_{10} \\
-b_{14} & b_{12} & -b_{1} & -b_{21} & b_{20} & -b_{18} & -b_{5} & -b_{9} \\
b_{13} & -b_{11} & b_{21} & -b_{1} & -b_{19} & b_{17} & b_{4} & b_{8} \\
b_{0} & b_{10} & -b_{20} & b_{19} & -b_{1} & -b_{16} & -b_{3} & -b_{7} \\
b_{10} & b_{0} & -b_{18} & b_{17} & -b_{16} & -b_{1} & -b_{2} & -b_{6} \\
b_{20} & -b_{18} & b_{0} & b_{15} & -b_{14} & b_{12} & -b_{9} & -b_{5} \\
-b_{19} & b_{17} & -b_{15} & b_{0} & b_{13} & -b_{11} & b_{8} & b_{4} \\
-b_{1} & -b_{16} & b_{14} & -b_{13} & b_{0} & b_{10} & -b_{7} & -b_{3} \\
-b_{16} & -b_{1} & b_{12} & -b_{11} & b_{10} & b_{0} & -b_{6} & -b_{2} \\
b_{3} & -b_{2} & -b_{9} & b_{8} & -b_{7} & b_{6} & b_{0} & b_{1} \\
-b_{7} & b_{6} & b_{5} & -b_{4} & b_{3} & -b_{2} & b_{1} & b_{0} \\
\end{matrix}
\end{array}
\right].
$
\end{adjustbox}
%MACIERZ KALUZY GOTOWA

\section{Synthesis of a rationalized algorithm for computing Kaluza numbers
product}

The direct multiplication of two Kaluza requires 1024 real multiplications and 992 real additions. We shall present the algorithm, which reduce arithmetical complexity to 512 real multiplications and 576 real additions.

At first, we rearrange the rows of the matrix in the following order \{1, 2, 3, 7, 5, 9, 4, 8, 6, 10, 11, 17, 13, 19, 15, 21, 12, 18, 14, 20, 16, 22, 23, 27, 25, 29, 24, 28, 26, 30, 31, 32\}. Next, we rearrange the columns of obtained matrix in the same manner. As a result, we obtain the following matrix:

Then we can rewrite expression (2.3) in following form:

\begin{equation}
\mathbf{Y}_{32\times1}=\mathbf{P}_{32}\breve{\mathbf{B}}_{32}\mathbf{P}_{32}\mathbf{X}_{32\times1} ,
\label{4}
\end{equation}
where
%MACIERZ P32

%%%%%%%%%%%%%%%%%%%%%%%%%%%%%%%%%%%%%%%%%%%%%%%%%%%%%%%%%%%
\[
\mathbf{P}_{32}=\left[
\begin{array}{ll}
\mathbf{P}_{16}^{(0,0)} & \mathbf{P}_{16}^{(1,0)} \\
\mathbf{P}_{16}^{(0,1)} & \mathbf{P}_{16}^{(1,1)}
\end{array}\right],
\]
\begin{center}
$\mathbf{P}_{16}^{(0,0)}$
\begin{adjustbox}{width=0.55\textwidth, margin= 0 5 0 0}
$
=\left[
\begin{array}{cc}
\begin{matrix}
1 & 0 & 0 & 0 & 0 & 0 & 0 & 0 \\
0 & 1 & 0 & 0 & 0 & 0 & 0 & 0 \\
0 & 0 & 1 & 0 & 0 & 0 & 0 & 0 \\
0 & 0 & 0 & 0 & 0 & 0 & 1 & 0 \\
0 & 0 & 0 & 0 & 1 & 0 & 0 & 0 \\
0 & 0 & 0 & 0 & 0 & 0 & 0 & 0 \\
0 & 0 & 0 & 1 & 0 & 0 & 0 & 0 \\
0 & 0 & 0 & 0 & 0 & 0 & 0 & 1 \\
0 & 0 & 0 & 0 & 0 & 1 & 0 & 0 \\
0 & 0 & 0 & 0 & 0 & 0 & 0 & 0 \\
0 & 0 & 0 & 0 & 0 & 0 & 0 & 0 \\
0 & 0 & 0 & 0 & 0 & 0 & 0 & 0 \\
0 & 0 & 0 & 0 & 0 & 0 & 0 & 0 \\
0 & 0 & 0 & 0 & 0 & 0 & 0 & 0 \\
0 & 0 & 0 & 0 & 0 & 0 & 0 & 0 \\
0 & 0 & 0 & 0 & 0 & 0 & 0 & 0 \\
\end{matrix}
&
\begin{matrix}
0 & 0 & 0 & 0 & 0 & 0 & 0 & 0 \\
0 & 0 & 0 & 0 & 0 & 0 & 0 & 0 \\
0 & 0 & 0 & 0 & 0 & 0 & 0 & 0 \\
0 & 0 & 0 & 0 & 0 & 0 & 0 & 0 \\
0 & 0 & 0 & 0 & 0 & 0 & 0 & 0 \\
1 & 0 & 0 & 0 & 0 & 0 & 0 & 0 \\
0 & 0 & 0 & 0 & 0 & 0 & 0 & 0 \\
0 & 0 & 0 & 0 & 0 & 0 & 0 & 0 \\
0 & 0 & 0 & 0 & 0 & 0 & 0 & 0 \\
0 & 1 & 0 & 0 & 0 & 0 & 0 & 0 \\
0 & 0 & 1 & 0 & 0 & 0 & 0 & 0 \\
0 & 0 & 0 & 0 & 0 & 0 & 0 & 0 \\
0 & 0 & 0 & 0 & 1 & 0 & 0 & 0 \\
0 & 0 & 0 & 0 & 0 & 0 & 0 & 0 \\
0 & 0 & 0 & 0 & 0 & 0 & 1 & 0 \\
0 & 0 & 0 & 0 & 0 & 0 & 0 & 0 \\
\end{matrix}
\end{array}
\right],
$
\end{adjustbox}%NW OK
\end{center}

\begin{center}
$\mathbf{P}_{16}^{(0,1)}$
\begin{adjustbox}{width=0.55\textwidth, margin= 0 0}
$
=\left[
\begin{array}{cc}
\begin{matrix}
0 & 0 & 0 & 0 & 0 & 0 & 0 & 0 \\
0 & 0 & 0 & 0 & 0 & 0 & 0 & 0 \\
0 & 0 & 0 & 0 & 0 & 0 & 0 & 0 \\
0 & 0 & 0 & 0 & 0 & 0 & 0 & 0 \\
0 & 0 & 0 & 0 & 0 & 0 & 0 & 0 \\
0 & 0 & 0 & 0 & 0 & 0 & 0 & 0 \\
0 & 0 & 0 & 0 & 0 & 0 & 0 & 0 \\
0 & 0 & 0 & 0 & 0 & 0 & 0 & 0 \\
0 & 0 & 0 & 0 & 0 & 0 & 0 & 0 \\
0 & 0 & 0 & 0 & 0 & 0 & 0 & 0 \\
0 & 0 & 0 & 0 & 0 & 0 & 0 & 0 \\
0 & 0 & 0 & 0 & 0 & 0 & 0 & 0 \\
0 & 0 & 0 & 0 & 0 & 0 & 0 & 0 \\
0 & 0 & 0 & 0 & 0 & 0 & 0 & 0 \\
0 & 0 & 0 & 0 & 0 & 0 & 0 & 0 \\
0 & 0 & 0 & 0 & 0 & 0 & 0 & 0 \\
\end{matrix}
&
\begin{matrix}
0 & 0 & 0 & 1 & 0 & 0 & 0 & 0 \\
0 & 0 & 0 & 0 & 0 & 0 & 0 & 0 \\
0 & 0 & 0 & 0 & 0 & 1 & 0 & 0 \\
0 & 0 & 0 & 0 & 0 & 0 & 0 & 0 \\
0 & 0 & 0 & 0 & 0 & 0 & 0 & 1 \\
0 & 0 & 0 & 0 & 0 & 0 & 0 & 0 \\
0 & 0 & 0 & 0 & 0 & 0 & 0 & 0 \\
0 & 0 & 0 & 0 & 0 & 0 & 0 & 0 \\
0 & 0 & 0 & 0 & 0 & 0 & 0 & 0 \\
0 & 0 & 0 & 0 & 0 & 0 & 0 & 0 \\
0 & 0 & 0 & 0 & 0 & 0 & 0 & 0 \\
0 & 0 & 0 & 0 & 0 & 0 & 0 & 0 \\
0 & 0 & 0 & 0 & 0 & 0 & 0 & 0 \\
0 & 0 & 0 & 0 & 0 & 0 & 0 & 0 \\
0 & 0 & 0 & 0 & 0 & 0 & 0 & 0 \\
0 & 0 & 0 & 0 & 0 & 0 & 0 & 0 \\
\end{matrix}
\end{array}
\right],
$
\end{adjustbox}%SW OK
\end{center}

\begin{center}
$\mathbf{P}_{16}^{(1,0)}$
\begin{adjustbox}{width=0.55\textwidth, margin= 0 0}
$
=\left[
\begin{array}{cc}
\begin{matrix}
0 & 0 & 0 & 0 & 0 & 0 & 0 & 0 \\
0 & 0 & 0 & 0 & 0 & 0 & 0 & 0 \\
0 & 0 & 0 & 0 & 0 & 0 & 0 & 0 \\
0 & 0 & 0 & 0 & 0 & 0 & 0 & 0 \\
0 & 0 & 0 & 0 & 0 & 0 & 0 & 0 \\
0 & 0 & 0 & 0 & 0 & 0 & 0 & 0 \\
0 & 0 & 0 & 0 & 0 & 0 & 0 & 0 \\
0 & 0 & 0 & 0 & 0 & 0 & 0 & 0 \\
0 & 0 & 0 & 0 & 0 & 0 & 0 & 0 \\
0 & 0 & 0 & 0 & 0 & 0 & 0 & 0 \\
0 & 0 & 0 & 0 & 0 & 0 & 0 & 0 \\
1 & 0 & 0 & 0 & 0 & 0 & 0 & 0 \\
0 & 0 & 0 & 0 & 0 & 0 & 0 & 0 \\
0 & 0 & 1 & 0 & 0 & 0 & 0 & 0 \\
0 & 0 & 0 & 0 & 0 & 0 & 0 & 0 \\
0 & 0 & 0 & 0 & 1 & 0 & 0 & 0 \\
\end{matrix}
&
\begin{matrix}
0 & 0 & 0 & 0 & 0 & 0 & 0 & 0 \\
0 & 0 & 0 & 0 & 0 & 0 & 0 & 0 \\
0 & 0 & 0 & 0 & 0 & 0 & 0 & 0 \\
0 & 0 & 0 & 0 & 0 & 0 & 0 & 0 \\
0 & 0 & 0 & 0 & 0 & 0 & 0 & 0 \\
0 & 0 & 0 & 0 & 0 & 0 & 0 & 0 \\
0 & 0 & 0 & 0 & 0 & 0 & 0 & 0 \\
0 & 0 & 0 & 0 & 0 & 0 & 0 & 0 \\
0 & 0 & 0 & 0 & 0 & 0 & 0 & 0 \\
0 & 0 & 0 & 0 & 0 & 0 & 0 & 0 \\
0 & 0 & 0 & 0 & 0 & 0 & 0 & 0 \\
0 & 0 & 0 & 0 & 0 & 0 & 0 & 0 \\
0 & 0 & 0 & 0 & 0 & 0 & 0 & 0 \\
0 & 0 & 0 & 0 & 0 & 0 & 0 & 0 \\
0 & 0 & 0 & 0 & 0 & 0 & 0 & 0 \\
0 & 0 & 0 & 0 & 0 & 0 & 0 & 0 \\
\end{matrix}
\end{array}
\right],
$
\end{adjustbox}%NE OK
\end{center}

\begin{center}
$\mathbf{P}_{16}^{(1,1)}$
\begin{adjustbox}{width=0.55\textwidth, margin= 0 0}
$
=\left[
\begin{array}{cc}
\begin{matrix}
0 & 0 & 0 & 0 & 0 & 0 & 0 & 0 \\
0 & 1 & 0 & 0 & 0 & 0 & 0 & 0 \\
0 & 0 & 0 & 0 & 0 & 0 & 0 & 0 \\
0 & 0 & 0 & 1 & 0 & 0 & 0 & 0 \\
0 & 0 & 0 & 0 & 0 & 0 & 0 & 0 \\
0 & 0 & 0 & 0 & 0 & 1 & 0 & 0 \\
0 & 0 & 0 & 0 & 0 & 0 & 1 & 0 \\
0 & 0 & 0 & 0 & 0 & 0 & 0 & 0 \\
0 & 0 & 0 & 0 & 0 & 0 & 0 & 0 \\
0 & 0 & 0 & 0 & 0 & 0 & 0 & 0 \\
0 & 0 & 0 & 0 & 0 & 0 & 0 & 1 \\
0 & 0 & 0 & 0 & 0 & 0 & 0 & 0 \\
0 & 0 & 0 & 0 & 0 & 0 & 0 & 0 \\
0 & 0 & 0 & 0 & 0 & 0 & 0 & 0 \\
0 & 0 & 0 & 0 & 0 & 0 & 0 & 0 \\
0 & 0 & 0 & 0 & 0 & 0 & 0 & 0 \\
\end{matrix}
&
\begin{matrix}
0 & 0 & 0 & 0 & 0 & 0 & 0 & 0 \\
0 & 0 & 0 & 0 & 0 & 0 & 0 & 0 \\
0 & 0 & 0 & 0 & 0 & 0 & 0 & 0 \\
0 & 0 & 0 & 0 & 0 & 0 & 0 & 0 \\
0 & 0 & 0 & 0 & 0 & 0 & 0 & 0 \\
0 & 0 & 0 & 0 & 0 & 0 & 0 & 0 \\
0 & 0 & 0 & 0 & 0 & 0 & 0 & 0 \\
0 & 0 & 1 & 0 & 0 & 0 & 0 & 0 \\
1 & 0 & 0 & 0 & 0 & 0 & 0 & 0 \\
0 & 0 & 0 & 0 & 1 & 0 & 0 & 0 \\
0 & 0 & 0 & 0 & 0 & 0 & 0 & 0 \\
0 & 0 & 0 & 1 & 0 & 0 & 0 & 0 \\
0 & 1 & 0 & 0 & 0 & 0 & 0 & 0 \\
0 & 0 & 0 & 0 & 0 & 1 & 0 & 0 \\
0 & 0 & 0 & 0 & 0 & 0 & 1 & 0 \\
0 & 0 & 0 & 0 & 0 & 0 & 0 & 1 \\
\end{matrix}
\end{array}
\right],
$
\end{adjustbox}%SE OK
\end{center}
%%%%%%%%%%%%%%%%%%%%%%%%%%%%%%%%%%%%%%%%%%%%%%%%%%%%%%%%%%%

and

% MACIERZ B32 Z AKCENTEM
\[
\breve{\mathbf{B}}_{32}=\left[
\begin{array}{ll}
\breve{\mathbf{B}}_{16}^{(0,0)} & \breve{\mathbf{B}}_{16}^{(1,0)} \\
\breve{\mathbf{B}}_{16}^{(0,1)} & \breve{\mathbf{B}}_{16}^{(1,1)}
\end{array}\right],
%\label{e1.5}
\]

%%TABELA przetasowana	
$\breve{\mathbf{B}}_{16}^{(0,0)}=$\\*
\begin{adjustbox}{width=\textwidth, margin=  3 15 0 5}
$
=\left[
\begin{array}{cc}
\begin{matrix}
b_{0} & b_{1} & b_{2} & -b_{6} & -b_{4} & b_{8} & -b_{3} & b_{7} \\
b_{1} & b_{0} & -b_{6} & b_{2} & b_{8} & -b_{4} & b_{7} & -b_{3} \\
b_{2} & b_{6} & b_{0} & -b_{1} & b_{11} & -b_{17} & b_{10} & -b_{16} \\
b_{6} & b_{2} & -b_{1} & b_{0} & -b_{17} & b_{11} & -b_{16} & b_{10} \\
b_{4} & b_{8} & b_{11} & -b_{17} & b_{0} & -b_{1} & -b_{13} & b_{19} \\
b_{8} & b_{4} & -b_{17} & b_{11} & -b_{1} & b_{0} & b_{19} & -b_{13} \\
b_{3} & b_{7} & b_{10} & -b_{16} & b_{13} & -b_{19} & b_{0} & -b_{1} \\
b_{7} & b_{3} & -b_{16} & b_{10} & -b_{19} & b_{13} & -b_{1} & b_{0} \\
b_{5} & b_{9} & b_{12} & -b_{18} & -b_{15} & b_{21} & -b_{14} & b_{20} \\
b_{9} & b_{5} & -b_{18} & b_{12} & b_{21} & -b_{15} & b_{20} & -b_{14} \\
b_{10} & b_{16} & b_{3} & -b_{7} & -b_{22} & b_{26} & -b_{2} & b_{6} \\
b_{16} & b_{10} & -b_{7} & b_{3} & b_{26} & -b_{22} & b_{6} & -b_{2} \\
b_{12} & b_{18} & b_{5} & -b_{9} & b_{24} & -b_{28} & b_{23} & -b_{27} \\
b_{18} & b_{12} & -b_{9} & b_{5} & -b_{28} & b_{24} & -b_{27} & b_{23} \\
b_{14} & b_{20} & b_{23} & -b_{27} & b_{25} & -b_{29} & b_{5} & -b_{9} \\
b_{20} & b_{14} & -b_{27} & b_{23} & -b_{29} & b_{25} & -b_{9} & b_{5} \\
\end{matrix}
&
\begin{matrix}
-b_{5} & b_{9} & b_{10} & b_{16} & b_{12} & b_{18} & -b_{14} & -b_{20} \\
b_{9} & -b_{5} & b_{16} & b_{10} & b_{18} & b_{12} & -b_{20} & -b_{14} \\
b_{12} & -b_{18} & -b_{3} & -b_{7} & -b_{5} & -b_{9} & -b_{23} & -b_{27} \\
-b_{18} & b_{12} & -b_{7} & -b_{3} & -b_{9} & -b_{5} & -b_{27} & -b_{23} \\
b_{15} & -b_{21} & b_{22} & b_{26} & -b_{24} & -b_{28} & b_{25} & b_{29} \\
-b_{21} & b_{15} & b_{26} & b_{22} & -b_{28} & -b_{24} & b_{29} & b_{25} \\
b_{14} & -b_{20} & -b_{2} & -b_{6} & -b_{23} & -b_{27} & -b_{5} & -b_{9} \\
-b_{20} & b_{14} & -b_{6} & -b_{2} & -b_{27} & -b_{23} & -b_{9} & -b_{5} \\
b_{0} & -b_{1} & b_{23} & b_{27} & -b_{2} & -b_{6} & b_{3} & b_{7} \\
-b_{1} & b_{0} & b_{27} & b_{23} & -b_{6} & -b_{2} & b_{7} & b_{3} \\
-b_{23} & b_{27} & b_{0} & b_{1} & b_{14} & b_{20} & -b_{12} & -b_{18} \\
b_{27} & -b_{23} & b_{1} & b_{0} & b_{20} & b_{14} & -b_{18} & -b_{12} \\
-b_{2} & b_{6} & -b_{14} & -b_{20} & b_{0} & b_{1} & b_{10} & b_{16} \\
b_{6} & -b_{2} & -b_{20} & -b_{14} & b_{1} & b_{0} & b_{16} & b_{10} \\
-b_{3} & b_{7} & -b_{12} & -b_{18} & b_{10} & b_{16} & b_{0} & b_{1} \\
b_{7} & -b_{3} & -b_{18} & -b_{12} & b_{16} & b_{10} & b_{1} & b_{0} \\
\end{matrix}
\end{array}
\right],
$
\end{adjustbox}%NW OK

$\breve{\mathbf{B}}_{16}^{(0,1)}=$\\*
\begin{adjustbox}{width=\textwidth, margin= 3 15 0 5}
$
=\left[
\begin{array}{cc}
\begin{matrix}
b_{11} & b_{17} & b_{4} & -b_{8} & -b_{2} & b_{6} & b_{22} & -b_{26} \\
b_{17} & b_{11} & -b_{8} & b_{4} & b_{6} & -b_{2} & -b_{26} & b_{22} \\
b_{13} & b_{19} & b_{22} & -b_{26} & -b_{3} & b_{7} & b_{4} & -b_{8} \\
b_{19} & b_{13} & -b_{26} & b_{22} & b_{7} & -b_{3} & -b_{8} & b_{4} \\
b_{15} & b_{21} & b_{24} & -b_{28} & b_{5} & -b_{9} & -b_{25} & b_{29} \\
b_{21} & b_{15} & -b_{28} & b_{24} & -b_{9} & b_{5} & b_{29} & -b_{25} \\
b_{22} & b_{26} & b_{13} & -b_{19} & b_{10} & -b_{16} & -b_{11} & b_{17} \\
b_{26} & b_{22} & -b_{19} & b_{13} & -b_{16} & b_{10} & b_{17} & -b_{11} \\
b_{24} & b_{28} & b_{15} & -b_{21} & -b_{12} & b_{18} & b_{30} & -b_{31} \\
b_{28} & b_{24} & -b_{21} & b_{15} & b_{18} & -b_{12} & -b_{31} & b_{30} \\
b_{23} & b_{27} & b_{14} & -b_{20} & -b_{30} & b_{31} & -b_{12} & b_{18} \\
b_{27} & b_{23} & -b_{20} & b_{14} & b_{31} & -b_{30} & b_{18} & -b_{12} \\
b_{25} & b_{29} & b_{30} & -b_{31} & -b_{14} & b_{20} & b_{15} & -b_{21} \\
b_{29} & b_{25} & -b_{31} & b_{30} & b_{20} & -b_{14} & -b_{21} & b_{15} \\
b_{30} & b_{31} & b_{25} & -b_{29} & b_{23} & -b_{27} & -b_{24} & b_{28} \\
b_{31} & b_{30} & -b_{29} & b_{25} & -b_{27} & b_{23} & b_{28} & -b_{24} \\
\end{matrix}
&
\begin{matrix}
-b_{24} & b_{28} & -b_{13} & -b_{19} & b_{15} & b_{21} & b_{30} & b_{31} \\
b_{28} & -b_{24} & -b_{19} & -b_{13} & b_{21} & b_{15} & b_{31} & b_{30} \\
-b_{25} & b_{29} & -b_{11} & -b_{17} & b_{30} & b_{31} & b_{15} & b_{21} \\
b_{29} & -b_{25} & -b_{17} & -b_{11} & b_{31} & b_{30} & b_{21} & b_{15} \\
-b_{4} & b_{8} & b_{30} & b_{31} & b_{11} & b_{17} & -b_{13} & -b_{19} \\
b_{8} & -b_{4} & b_{31} & b_{30} & b_{17} & b_{11} & -b_{19} & -b_{13} \\
b_{30} & -b_{31} & b_{4} & b_{8} & -b_{25} & -b_{29} & b_{24} & b_{28} \\
-b_{31} & b_{30} & b_{8} & b_{4} & -b_{29} & -b_{25} & b_{28} & b_{24} \\
b_{11} & -b_{17} & -b_{25} & -b_{29} & -b_{4} & -b_{8} & -b_{22} & -b_{26} \\
-b_{17} & b_{11} & -b_{29} & -b_{25} & -b_{8} & -b_{4} & -b_{26} & -b_{22} \\
b_{10} & -b_{16} & b_{5} & b_{9} & -b_{3} & -b_{7} & b_{2} & b_{6} \\
-b_{16} & b_{10} & b_{9} & b_{5} & -b_{7} & -b_{3} & b_{6} & b_{2} \\
b_{13} & -b_{19} & -b_{24} & -b_{28} & -b_{22} & -b_{26} & -b_{4} & -b_{8} \\
-b_{19} & b_{13} & -b_{28} & -b_{24} & -b_{26} & -b_{22} & -b_{8} & -b_{4} \\
-b_{22} & b_{26} & b_{15} & b_{21} & b_{13} & b_{19} & -b_{11} & -b_{17} \\
b_{26} & -b_{22} & b_{21} & b_{15} & b_{19} & b_{13} & -b_{17} & -b_{11} \\
\end{matrix}
\end{array}
\right],
$
\end{adjustbox}%SW OK

$\breve{\mathbf{B}}_{16}^{(1,0)}=$\\*
\begin{adjustbox}{width=\textwidth, margin= 3 15 0 5}
$
=\left[
\begin{array}{cc}
\begin{matrix}
b_{11} & b_{17} & -b_{13} & -b_{19} & -b_{15} & -b_{21} & -b_{22} & b_{26} \\
b_{17} & b_{11} & -b_{19} & -b_{13} & -b_{21} & -b_{15} & b_{26} & -b_{22} \\
-b_{4} & -b_{8} & -b_{22} & -b_{26} & -b_{24} & -b_{28} & -b_{13} & b_{19} \\
-b_{8} & -b_{4} & -b_{26} & -b_{22} & -b_{28} & -b_{24} & b_{19} & -b_{13} \\
-b_{2} & -b_{6} & b_{3} & b_{7} & -b_{5} & -b_{9} & b_{10} & -b_{16} \\
-b_{6} & -b_{2} & b_{7} & b_{3} & -b_{9} & -b_{5} & -b_{16} & b_{10} \\
-b_{22} & -b_{26} & -b_{4} & -b_{8} & -b_{25} & -b_{29} & -b_{11} & b_{17} \\
-b_{26} & -b_{22} & -b_{8} & -b_{4} & -b_{29} & -b_{25} & b_{17} & -b_{11} \\
b_{24} & b_{28} & -b_{25} & -b_{29} & b_{4} & b_{8} & -b_{30} & b_{31} \\
b_{28} & b_{24} & -b_{29} & -b_{25} & b_{8} & b_{4} & b_{31} & -b_{30} \\
b_{13} & b_{19} & -b_{11} & -b_{17} & -b_{30} & -b_{31} & -b_{4} & b_{8} \\
b_{19} & b_{13} & -b_{17} & -b_{11} & -b_{31} & -b_{30} & b_{8} & -b_{4} \\
-b_{15} & -b_{21} & -b_{30} & -b_{31} & b_{11} & b_{17} & -b_{25} & b_{29} \\
-b_{21} & -b_{15} & -b_{31} & -b_{30} & b_{17} & b_{11} & b_{29} & -b_{25} \\
-b_{30} & -b_{31} & -b_{15} & -b_{21} & b_{13} & b_{19} & -b_{24} & b_{28} \\
-b_{31} & -b_{30} & -b_{21} & -b_{15} & b_{19} & b_{13} & b_{28} & -b_{24} \\
\end{matrix}
&
\begin{matrix}
-b_{24} & b_{28} & -b_{23} & b_{27} & b_{25} & -b_{29} & -b_{30} & -b_{31} \\
b_{28} & -b_{24} & b_{27} & -b_{23} & -b_{29} & b_{25} & -b_{31} & -b_{30} \\
-b_{15} & b_{21} & -b_{14} & b_{20} & -b_{30} & b_{31} & b_{25} & b_{29} \\
b_{21} & -b_{15} & b_{20} & -b_{14} & b_{31} & -b_{30} & b_{29} & b_{25} \\
-b_{12} & b_{18} & b_{30} & -b_{31} & b_{14} & -b_{20} & -b_{23} & -b_{27} \\
b_{18} & -b_{12} & -b_{31} & b_{30} & -b_{20} & b_{14} & -b_{27} & -b_{23} \\
-b_{30} & b_{31} & -b_{12} & b_{18} & -b_{15} & b_{21} & b_{24} & b_{28} \\
b_{31} & -b_{30} & b_{18} & -b_{12} & b_{21} & -b_{15} & b_{28} & b_{24} \\
b_{11} & -b_{17} & b_{10} & -b_{16} & -b_{13} & b_{19} & b_{22} & b_{26} \\
-b_{17} & b_{11} & -b_{16} & b_{10} & b_{19} & -b_{13} & b_{26} & b_{22} \\
-b_{25} & b_{29} & -b_{5} & b_{9} & b_{24} & -b_{28} & -b_{15} & -b_{21} \\
b_{29} & -b_{25} & b_{9} & -b_{5} & -b_{28} & b_{24} & -b_{21} & -b_{15} \\
b_{4} & -b_{8} & b_{3} & -b_{7} & b_{22} & -b_{26} & -b_{13} & -b_{19} \\
-b_{8} & b_{4} & -b_{7} & b_{3} & -b_{26} & b_{22} & -b_{19} & -b_{13} \\
b_{22} & -b_{26} & b_{2} & -b_{6} & b_{4} & -b_{8} & -b_{11} & -b_{17} \\
-b_{26} & b_{22} & -b_{6} & b_{2} & -b_{8} & b_{4} & -b_{17} & -b_{11} \\
\end{matrix}
\end{array}
\right],
$
\end{adjustbox}%NE OK

$\breve{\mathbf{B}}_{16}^{(1,1)}=$\\*
\begin{adjustbox}{width=\textwidth, margin= 3 15 0 5}
$
=\left[
\begin{array}{cc}
\begin{matrix}
b_{0} & b_{1} & b_{10} & b_{16} & -b_{12} & -b_{18} & b_{3} & -b_{7} \\
b_{1} & b_{0} & b_{16} & b_{10} & -b_{18} & -b_{12} & -b_{7} & b_{3} \\
b_{10} & b_{16} & b_{0} & b_{1} & -b_{14} & -b_{20} & b_{2} & -b_{6} \\
b_{16} & b_{10} & b_{1} & b_{0} & -b_{20} & -b_{14} & -b_{6} & b_{2} \\
-b_{12} & -b_{18} & b_{14} & b_{20} & b_{0} & b_{1} & b_{23} & -b_{27} \\
-b_{18} & -b_{12} & b_{20} & b_{14} & b_{1} & b_{0} & -b_{27} & b_{23} \\
-b_{3} & -b_{7} & b_{2} & b_{6} & -b_{23} & -b_{27} & b_{0} & -b_{1} \\
-b_{7} & -b_{3} & b_{6} & b_{2} & -b_{27} & -b_{23} & -b_{1} & b_{0} \\
b_{5} & b_{9} & b_{23} & b_{27} & b_{2} & b_{6} & b_{14} & -b_{20} \\
b_{9} & b_{5} & b_{27} & b_{23} & b_{6} & b_{2} & -b_{20} & b_{14} \\
b_{25} & b_{29} & -b_{24} & -b_{28} & b_{22} & b_{26} & -b_{15} & b_{21} \\
b_{29} & b_{25} & -b_{28} & -b_{24} & b_{26} & b_{22} & b_{21} & -b_{15} \\
b_{23} & b_{27} & b_{5} & b_{9} & b_{3} & b_{7} & b_{12} & -b_{18} \\
b_{27} & b_{23} & b_{9} & b_{5} & b_{7} & b_{3} & -b_{18} & b_{12} \\
-b_{14} & -b_{20} & b_{12} & b_{18} & b_{10} & b_{16} & b_{5} & -b_{9} \\
-b_{20} & -b_{14} & b_{18} & b_{12} & b_{16} & b_{10} & -b_{9} & b_{5} \\
\end{matrix}
&
\begin{matrix}
-b_{5} & b_{9} & b_{25} & -b_{29} & -b_{23} & b_{27} & b_{14} & b_{20} \\
b_{9} & -b_{5} & -b_{29} & b_{25} & b_{27} & -b_{23} & b_{20} & b_{14} \\
-b_{23} & b_{27} & b_{24} & -b_{28} & -b_{5} & b_{9} & b_{12} & b_{18} \\
b_{27} & -b_{23} & -b_{28} & b_{24} & b_{9} & -b_{5} & b_{18} & b_{12} \\
b_{2} & -b_{6} & -b_{22} & b_{26} & -b_{3} & b_{7} & b_{10} & b_{16} \\
-b_{6} & b_{2} & b_{26} & -b_{22} & b_{7} & -b_{3} & b_{16} & b_{10} \\
-b_{14} & b_{20} & b_{15} & -b_{21} & b_{12} & -b_{18} & -b_{5} & -b_{9} \\
b_{20} & -b_{14} & -b_{21} & b_{15} & -b_{18} & b_{12} & -b_{9} & -b_{5} \\
b_{0} & -b_{1} & -b_{13} & b_{19} & b_{10} & -b_{16} & -b_{3} & -b_{7} \\
-b_{1} & b_{0} & b_{19} & -b_{13} & -b_{16} & b_{10} & -b_{7} & -b_{3} \\
b_{13} & -b_{19} & b_{0} & -b_{1} & -b_{11} & b_{17} & b_{4} & b_{8} \\
-b_{19} & b_{13} & -b_{1} & b_{0} & b_{17} & -b_{11} & b_{8} & b_{4} \\
b_{10} & -b_{16} & -b_{11} & b_{17} & b_{0} & -b_{1} & -b_{2} & -b_{6} \\
b_{16} & b_{10} & b_{17} & -b_{11} & -b_{1} & b_{0} & -b_{6} & -b_{2} \\
b_{3} & -b_{7} & -b_{4} & b_{8} & -b_{2} & b_{6} & b_{0} & b_{1} \\
-b_{7} & b_{3} & b_{8} & -b_{4} & b_{6} & -b_{2} & b_{1} & b_{0} \\
\end{matrix}
\end{array}
\right],
$
\end{adjustbox}%SE OK
%Macierz przetasowana gotowa

If we interpret the resulting matrix as a block-matrix, it is easy to see that each block of this matrix (or more precisely, each ($2\times2$)-submatrix) is bisymmetric, i.e. has the properties of pair wise symmetry about both of its main diagonals. There is an effective method of factorization of this type matrices, which allows during the calculation of the vector-matrix products halve the number of multiplications, that is to perform only two, not four multiplications at the expense of triple increasing the number of additions (from two to six) \cite{25,26}:
\begin{center}
\begin{equation}
\begin{adjustbox}{width=0.8\textwidth, margin=0 10 0 0}
$
\left[
\begin{array}{ll}
a & b \\
b & a
\end{array}
\right]
\left[
\begin{array}{c}
x_{0} \\
x_{1}
\end{array}
\right]
=\left[
\begin{array}{cc}
1 & 1 \\
1 & -1
\end{array}
\right]
\left[
\begin{array}{cc}
\frac{1}{2}(a+b) & 0 \\
0 & \frac{1}{2}(a-b)
\end{array}
\right]
\left[
\begin{array}{cc}
1 & 1 \\
1 & -1
\end{array}
\right]
\left[
\begin{array}{c}
x_{0} \\
x_{1}
\end{array}
\right] ,
$
\end{adjustbox}
\label{5}
\end{equation}
\end{center}
where $\left[
\begin{array}{cc}
1 & 1 \\
1 & -1
\end{array}
\right]=\mathbf{H}_{2}$ is the ($2\times2$) Hadamard matrix.

Fig. 1. shows a data flow diagram of the rationalized algorithm for ($2\times2$) bisymmetric
matrix-vector multiplication in according to (3.2). The circles in this figure show the operation
of multiplication by a value inscribed inside a circle. In turn, the rectangles indicate the
matrix-vector multiplications with matrices inscribed inside rectangles. In this paper the data flow diagrams are oriented from left to right. Straight lines in the figure denote the operation of data transfer. We use the solid lines without any arrows, so as not to clutter up the presented diagrams.

Now it is easy to show that using the above method of ($2\times2$)-bisymmetric matrix factorization
we can construct an efficient algorithm for multiplying Kaluza numbers. In such algorithm the number of real multiplications will be reduced by half  compared with naive method of calculations. However, if we directly use an expansion (3.2) for
each of the submatrices of matrix $\breve{\mathbf{B}}_{32}$, the number of real additions will increases dramatically.
In this case, the total value of computation complexity becomes even greater than in
the case of naive method of computations. However, if we utilize the fact that the
multiplication of the ($2\times2$) -Hadamard matrix on the corresponding subvector (the fig. 1 is
positioned before of multiplications) is a common operation for all submatrices disposed in
a same vertical of matrix $\breve{\mathbf{B}}_{32}$, and a similar operation (in the fig.1, it follows by multiplications) is a common operation for all submatrices disposed in a same horizontal of matrix $\breve{\mathbf{B}}_{32}$, the number of real additions can be significantly reduced. Consider the synthesis of an efficient algorithm for multiplying Kaluza numbers in more detail.

%\DeclareGraphicsExtensions{.pdf,.png,.jpg}
%RYSUNEK
\begin{figure}[ht]
\centering
  \includegraphics[width=0.3\textwidth]{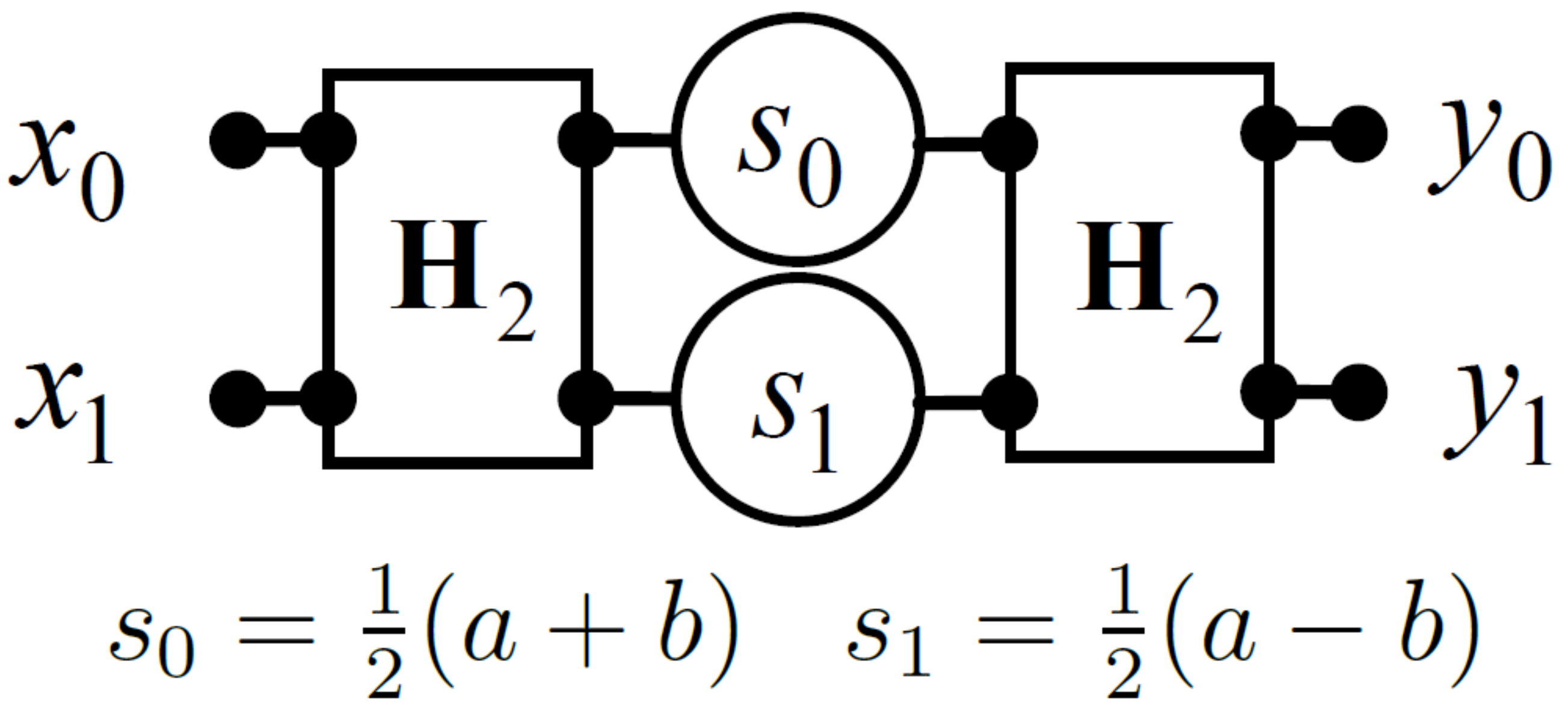}
  \caption{Data flow diagram of the rationalized algorithm for ($2\times2$)-bisymmetric matrix-vector
multiplication in according to (3.2)}
  \label{fig1}
\end{figure}

Let us first introduce some matrices

\[
\mathbf{W}_{16}=(\mathbf{I}_{16} \otimes \mathbf{H}_{2}),
 \mathbf{P}_{512\times32}=\mathbf{I}_{16}\otimes(\mathbf{I}_{16\times1}\otimes\mathbf{I}_{2}),
\]\[
\mathbf{D}_{512}=\bigoplus_{k=0}^{15}\mathbf{D}_{32}^{(k)},
 \mathbf{A}_{32\times512}=(\mathbf{I}_{16\times1}\otimes\mathbf{I}_{32}),
\]
where $\mathbf{I}_{N}$ is an identity $N\times{N}$ matrix, signs ``$\otimes$'' and ``$\oplus$'' denote tensor product and direct sum of two matrices respectively, and $\mathbf{I}_{N\times{M}}$  is an integer matrix consisting of all 1s [26],

\[
\mathbf{D}_{32}^{(k)}=diag(s_{0}^{(k)},s_{1}^{(k)},\dots,s_{31}^{(k)}), k=0,1,\dots,15,
\]
where

% S=C
\begin{center}
\begin{adjustbox}{width=0.54\textwidth, margin=0 0 0 10}
$
\begin{array}{cccc}
\renewcommand*{\arraystretch}{1.5}
\begin{matrix}
s_{0}^{(0)}=c_{0}, & s_{1}^{(0)}=c_{1}, & s_{2}^{(0)}=c_{2}, & s_{3}^{(0)}=c_{3}, \\
s_{4}^{(0)}=c_{4}, & s_{5}^{(0)}=c_{5}, & s_{6}^{(0)}=c_{6}, & s_{7}^{(0)}=c_{7}, \\
s_{8}^{(0)}=c_{8}, & s_{9}^{(0)}=c_{9}, & s_{10}^{(0)}=c_{10}, & s_{11}^{(0)}=c_{11}, \\
s_{12}^{(0)}=c_{12}, & s_{13}^{(0)}=c_{13}, & s_{14}^{(0)}=c_{14}, & s_{15}^{(0)}=c_{15}, \\
s_{16}^{(0)}=c_{16}, & s_{17}^{(0)}=c_{17}, & s_{18}^{(0)}=c_{18}, & s_{19}^{(0)}=c_{19}, \\
s_{20}^{(0)}=c_{20}, & s_{21}^{(0)}=c_{21}, & s_{22}^{(0)}=c_{22}, & s_{23}^{(0)}=c_{23}, \\
s_{24}^{(0)}=c_{24}, & s_{25}^{(0)}=c_{25}, & s_{26}^{(0)}=c_{26}, & s_{27}^{(0)}=c_{27}, \\
s_{28}^{(0)}=c_{28}, & s_{29}^{(0)}=c_{29}, & s_{30}^{(0)}=c_{30}, & s_{31}^{(0)}=c_{31}, \\
\end{matrix}
\end{array}
$
\end{adjustbox}
\end{center}

\begin{center}
\begin{adjustbox}{width=0.57\textwidth, margin=0 0 0 0}
$
\begin{array}{cccc}
\renewcommand*{\arraystretch}{1.5}
\begin{matrix}
s_{0}^{(1)}=c_{3}, & s_{1}^{(1)}=c_{2}, & s_{2}^{(1)}=c_{1}, & s_{3}^{(1)}=c_{0}, \\
s_{4}^{(1)}=c_{17}, & s_{5}^{(1)}=c_{16}, & s_{6}^{(1)}=c_{11}, & s_{7}^{(1)}=c_{10}, \\
s_{8}^{(1)}=c_{13}, & s_{9}^{(1)}=c_{12}, & s_{10}^{(1)}=c_{7}, & s_{11}^{(1)}=c_{6}, \\
s_{12}^{(1)}=c_{9}, & s_{13}^{(1)}=c_{8}, & s_{14}^{(1)}=c_{27}, & s_{15}^{(1)}=c_{26}, \\
s_{16}^{(1)}=c_{5}, & s_{17}^{(1)}=c_{4}, & s_{18}^{(1)}=-c_{22}, & s_{19}^{(1)}=-c_{23}, \\
s_{20}^{(1)}=c_{25}, & s_{21}^{(1)}=c_{24}, & s_{22}^{(1)}=c_{19}, & s_{23}^{(1)}=c_{18}, \\
s_{24}^{(1)}=c_{21}, & s_{25}^{(1)}=c_{20}, & s_{26}^{(1)}=c_{15}, & s_{27}^{(1)}=c_{14}, \\
s_{28}^{(1)}=c_{31}, & s_{29}^{(1)}=c_{30}, & s_{30}^{(1)}=c_{29}, & s_{31}^{(1)}=c_{28}, \\
\end{matrix}
\end{array}
$
\end{adjustbox}
\end{center}

\begin{center}
\begin{adjustbox}{width=0.57\textwidth, margin=0 0 0 0}
$
\begin{array}{cccc}
\renewcommand*{\arraystretch}{1.5}
\begin{matrix}
s_{0}^{(2)}=-c_{5}, & 	s_{1}^{(2)}=-c_{4}, & 	s_{2}^{(2)}=c_{17}, & 	s_{3}^{(2)}=c_{16}, \\
s_{4}^{(2)}=c_{1}, & 	s_{5}^{(2)}=c_{0}, & 	s_{6}^{(2)}=c_{19}, & 	  s_{7}^{(2)} =c_{18}, \\
s_{8}^{(2)}=-c_{21}, & 	s_{9}^{(2)}=-c_{20}, & 	s_{10}^{(2)}=-c_{23}, & 	s_{11}^{(2)}=-c_{22}, \\ 	  s_{12}^{(2)}=c_{25}, & 	s_{13}^{(2)}=c_{24}, & 	s_{14}^{(2)}=c_{29}, & 	s_{15}^{(2)}=c_{28}, \\
s_{16}^{(2)}=-c_{3}, & 	s_{17}^{(2)}=-c_{2}, & 	s_{18}^{(2)}=-c_{7}, & 	s_{19}^{(2)}=-c_{6}, \\ 	s_{20}^{(2)}=c_{9}, & 	s_{21}^{(2)}=c_{8}, & 	s_{22}^{(2)}=c_{11}, & 	s_{23}^{(2)}=c_{10}, \\
s_{24}^{(2)}=-c_{13}, & 	s_{25}^{(2)}=-c_{12}, & 	s_{26}^{(2)}=-c_{31}, & 	s_{27}^{(2)}=-c_{30}, \\
s_{28}^{(2)}=-c_{15}, & 	s_{29}^{(2)}=-c_{14}, & 	s_{30}^{(2)}=c_{27}, & 	s_{31}^{(2)}=c_{26}, \\
 \end{matrix}
\end{array}
$
\end{adjustbox}
\end{center}

\begin{center}
\begin{adjustbox}{width=0.57\textwidth, margin=0 0 0 0}
$
\begin{array}{cccc}
\renewcommand*{\arraystretch}{1.5}
\begin{matrix}
s_{0}^{(3)}=-c_{7}, & 	s_{1}^{(3)}=-c_{6}, & 	s_{2}^{(3)}=c_{11}, & 	s_{3}^{(3)}=c_{10}, \\
s_{4}^{(3)}=-c_{19}, & 	s_{5}^{(3)}=-c_{18}, & 	s_{6}^{(3)}=c_{1}, & 	  s_{7}^{(3)}=c_{0}, \\
s_{8}^{(3)}=-c_{15}, & 	s_{9}^{(3)}=-c_{14}, & 	s_{10}^{(3)}=-c_{3}, & 	s_{11}^{(3)}=-c_{2}, \\
s_{12}^{(3)}=c_{27}, & 	s_{13}^{(3)}=c_{26}, & 	s_{14}^{(3)}=c_{9}, & 	s_{15}^{(3)}=c_{8}, \\
s_{16}^{(3)}=c_{23}, & 	s_{17}^{(3)}=c_{22}, & 	s_{18}^{(3)}=c_{5}, & 	s_{19}^{(3)}=c_{4}, \\
s_{20}^{(3)}=-c_{29}, & 	s_{21}^{(3)}=-c_{28}, & 	s_{22}^{(3)}=-c_{17}, & 	s_{23}^{(3)} =-c_{16}, \\
s_{24}^{(3)}=c_{31}, & 	s_{25}^{(3)}=c_{30}, & 	s_{26}^{(3)}=-c_{13}, & 	s_{27}^{(3)}=-c_{12}, \\
s_{28}^{(3)}=c_{21}, & 	s_{29}^{(3)}=c_{20}, & 	s_{30}^{(3)}=-c_{25}, & 	s_{31}^{(3)}=-c_{24}, \\

\end{matrix}
\end{array}
$
\end{adjustbox}
\end{center}

\begin{center}
\begin{adjustbox}{width=0.57\textwidth, margin=0 0 0 0}
$
\begin{array}{cccc}
\renewcommand*{\arraystretch}{1.5}
\begin{matrix}

s_{0}^{(4)}=-c_{9}, & 	s_{1}^{(4)}=-c_{8}, & 	s_{2}^{(4)}=c_{13}, & 	s_{3}^{(4)}=c_{12}, \\ 	s_{4}^{(4)}=c_{21}, & 	s_{5}^{(4)}=c_{20}, & 	s_{6}^{(4)}=c_{15}, & 	s_{7}^{(4)}=c_{14}, \\
s_{8}^{(4)}=c_{1}, & 	s_{9}^{(4)}=c_{0}, & 	s_{10}^{(4)}=-c_{27}, & 	s_{11}^{(4)}=-c_{26}, \\
s_{12}^{(4)}=-c_{3}, & 	s_{13}^{(4)}=-c_{2}, & 	s_{14}^{(4)}=-c_{7}, & 	s_{15}^{(4)}=-c_{6}, \\
s_{16}^{(4)}=-c_{25}, & 	s_{17}^{(4)}=-c_{24}, & 	s_{18}^{(4)}=-c_{29}, & 	s_{19}^{(4)}=-c_{28}, \\
s_{20}^{(4)}=-c_{5}, & 	s_{21}^{(4)}=-c_{4}, & 	s_{22}^{(4)}=c_{31}, & 	s_{23}^{(4)}=c_{30}, \\
s_{24}^{(4)}=c_{17}, & 	s_{25}^{(4)}=c_{16}, & 	s_{26}^{(4)}=c_{11}, & 	s_{27}^{(4)}=c_{10}, \\ 	s_{28}^{(4)}=c_{19}, & 	s_{29}^{(4)}=c_{18}, & 	s_{30}^{(4)}=-c_{23}, & 	s_{31}^{(4)}=-c_{22}, \\
\end{matrix}
\end{array}
$
\end{adjustbox}
\end{center}

\begin{center}
\begin{adjustbox}{width=0.57\textwidth, margin=0 0 0 0}
$
\begin{array}{cccc}
\renewcommand*{\arraystretch}{1.5}
\begin{matrix}
s_{0}^{(5)}=c_{10}, & 	s_{1}^{(5)}=c_{11}, & 	s_{2}^{(5)}=-c_{6}, & 	s_{3}^{(5)}=-c_{7}, \\ 	s_{4}^{(5)}=c_{22}, & 	s_{5}^{(5)}=c_{23}, & 	s_{6}^{(5)}=-c_{2}, & 	s_{7}^{(5)}=-c_{3}, \\
s_{8}^{(5)}=c_{26}, & 	s_{9}^{(5)}=c_{27}, & 	s_{10}^{(5)}=c_{0}, & 	s_{11}^{(5)}=c_{1}, \\
s_{12}^{(5)}=-c_{14}, & 	s_{13}^{(5)}=-c_{15}, & 	s_{14}^{(5)}=-c_{12}, & 	s_{15}^{(5)}=-c_{13}, \\
s_{16}^{(5)}=-c_{18}, & 	s_{17}^{(5)}=-c_{19}, & 	s_{18}^{(5)}=-c_{16}, & 	s_{19}^{(5)}=-c_{17}, \\ 	s_{20}^{(5)}=c_{30}, & 	s_{21}^{(5)}=c_{31}, & 	s_{22}^{(5)}=c_{4}, & 	s_{23}^{(5)}=c_{5}, \\
s_{24}^{(5)}=-c_{28}, & 	s_{25}^{(5)}=-c_{29}, & 	s_{26}^{(5)}=c_{8}, & 	s_{27}^{(5)}=c_{9}, \\
s_{28}^{(5)}=-c_{24}, & 	s_{29}^{(5)}=-c_{25}, & 	s_{30}^{(5)}=c_{20}, & 	s_{31}^{(5)}=c_{21}, \\
\end{matrix}
\end{array}
$
\end{adjustbox}
\end{center}

\begin{center}
\begin{adjustbox}{width=0.57\textwidth, margin=0 0 0 0}
$
\begin{array}{cccc}
\renewcommand*{\arraystretch}{1.5}
\begin{matrix}
s_{0}^{(6)}=c_{12}, & 	s_{1}^{(6)}=c_{13}, & 	s_{2}^{(6)}=-c_{8}, & 	s_{3}^{(6)}=-c_{9}, \\
s_{4}^{(6)}=-c_{24}, & 	s_{5}^{(6)}=-c_{25}, & 	s_{6}^{(6)}=-c_{26}, & 	      s_{7}^{(6)}=-c_{27}, \\
s_{8}^{(6)}=-c_{2}, & 	s_{9}^{(6)}=-c_{3}, & 	s_{10}^{(6)}=c_{14}, & 	s_{11}^{(6)}=c_{15}, \\ 	s_{12}^{(6)}=c_{0}, & 	s_{13}^{(6)}=c_{1}, & 	s_{14}^{(6)}=c_{10}, & 	s_{15}^{(6)}=c_{11}, \\
s_{16}^{(6)}=c_{20}, & 	s_{17}^{(6)}=c_{21}, & 	s_{18}^{(6)}=c_{30}, & 	s_{19}^{(6)}=c_{31}, \\ 	s_{20}^{(6)}=c_{16}, & 	s_{21}^{(6)}=c_{17}, & 	s_{22}^{(6)}=-c_{28}, & 	s_{23}^{(6)}=-c_{29}, \\
s_{24}^{(6)}=-c_{4}, & 	s_{25}^{(6)}=-c_{5}, & 	s_{26}^{(6)}=-c_{6}, & 	s_{27}^{(6)}=-c_{7}, \\
s_{28}^{(6)}=-c_{22}, & 	s_{29}^{(6)}=-c_{23}, & 	s_{30}^{(6)}=c_{18}, & 	s_{31}^{(6)}=c_{19}, \\
\end{matrix}
\end{array}
$
\end{adjustbox}
\end{center}

\begin{center}
\begin{adjustbox}{width=0.57\textwidth, margin=0 0 0 0}
$
\begin{array}{cccc}
\renewcommand*{\arraystretch}{1.5}
\begin{matrix}
s_{0}^{(7)}=-c_{14}, & 	s_{1}^{(7)}=-c_{15}, & 	s_{2}^{(7)}=-c_{26}, & 	s_{3}^{(7)}=-c_{27}, \\ 	s_{4}^{(7)}=c_{28}, & 	s_{5}^{(7)}=c_{29}, & 	s_{6}^{(7)}=-c_{8}, & 	s_{7}^{(7)}=-c_{9}, \\
s_{8}^{(7)}=c_{6}, & 	s_{9}^{(7)}=c_{7}, & 	s_{10}^{(7)}=-c_{12}, & 	s_{11}^{(7)}=-c_{13}, \\	s_{12}^{(7)}=c_{10}, & 	s_{13}^{(7)}=c_{11}, & 	s_{14}^{(7)}=c_{0}, & 	s_{15}^{(7)}=c_{1}, \\
s_{16}^{(7)}=c_{30}, & 	s_{17}^{(7)}=c_{31}, & 	s_{18}^{(7)}=c_{20}, & 	s_{19}^{(7)}=c_{21}, \\
s_{20}^{(7)}=-c_{18}, & 	s_{21}^{(7)}=-c_{19}, & 	s_{22}^{(7)}=c_{24}, & 	s_{23}^{(7)}=c_{25}, \\
s_{24}^{(7)}=-c_{22}, & 	s_{25}^{(7)}=-c_{23}, & 	s_{26}^{(7)}=c_{2}, & 	s_{27}^{(7)}=c_{3}, \\
s_{28}^{(7)}=-c_{4}, & 	s_{29}^{(7)}=-c_{5}, & 	s_{30}^{(7)}=-c_{16}, & 	s_{31}^{(7)}=-c_{17}, \\
\end{matrix}
\end{array}
$
\end{adjustbox}
\end{center}

\begin{center}
\begin{adjustbox}{width=0.57\textwidth, margin=0 0 0 0}
$
\begin{array}{cccc}
\renewcommand*{\arraystretch}{1.5}
\begin{matrix}
s_{0}^{(8)}=c_{16}, & 	s_{1}^{(8)}=c_{17}, & 	s_{2}^{(8)}=-c_{4}, & 	s_{3}^{(8)}=-c_{5}, \\
s_{4}^{(8)}=-c_{2}, & 	s_{5}^{(8)}=-c_{3}, & 	s_{6}^{(8)}=-c_{22}, & 	s_{7}^{(8)}=-c_{23}, \\
s_{8}^{(8)}=c_{24}, & 	s_{9}^{(8)}=c_{25}, & 	s_{10}^{(8)}=c_{18}, & 	s_{11}^{(8)}=c_{19}, \\
s_{12}^{(8)}=-c_{20}, & 	s_{13}^{(8)}=-c_{21}, & 	s_{14}^{(8)}=-c_{30}, & 	s_{15}^{(8)}=-c_{31}, \\
s_{16}^{(8)}=c_{0}, & 	s_{17}^{(8)}=c_{1}, & 	s_{18}^{(8)}=c_{10}, & 	s_{19}^{(8)}=c_{11}, \\
s_{20}^{(8)}=-c_{12}, & 	s_{21}^{(8)}=-c_{13}, & 	s_{22}^{(8)}=-c_{6}, & 	s_{23}^{(8)}=-c_{7}, \\
s_{24}^{(8)}=c_{8}, & 	s_{25}^{(8)}=c_{9}, & 	s_{26}^{(8)}=c_{28}, & 	s_{27}^{(8)}=c_{29}, \\	s_{28}^{(8)}=c_{26}, & 	s_{29}^{(8)}=c_{27}, & 	s_{30}^{(8)}=-c_{14}, & 	s_{31}^{(8)}=-c_{15}, \\
\end{matrix}
\end{array}
$
\end{adjustbox}
\end{center}

\begin{center}
\begin{adjustbox}{width=0.57\textwidth, margin=0 0 0 0}
$
\begin{array}{cccc}
\renewcommand*{\arraystretch}{1.5}
\begin{matrix}
s_{0}^{(9)}=-c_{18}, & 	s_{1}^{(9)}=-c_{19}, & 	s_{2}^{(9)}=-c_{22}, & 	s_{3}^{(9)}=-c_{23}, \\
s_{4}^{(9)}=c_{6}, & 	s_{5}^{(9)}=c_{7}, & 	s_{6}^{(9)}=-c_{4}, & 	s_{7}^{(9)}=-c_{5}, \\
s_{8}^{(9)}=-c_{28}, & 	s_{9}^{(9)}=-c_{29}, & 	s_{10}^{(9)}=-c_{16}, & 	s_{11}^{(9)}=-c_{17}, \\
s_{12}^{(9)}=-c_{30}, & 	s_{13}^{(9)}=-c_{31}, & 	s_{14}^{(9)}=-c_{20}, & 	s_{15}^{(9)}=-c_{21}, \\
s_{16}^{(9)}=c_{10}, & 	s_{17}^{(9)}=c_{11}, & 	s_{18}^{(9)}=c_{0}, & 	s_{19}^{(9)}=c_{1}, \\
s_{20}^{(9)}=c_{14}, & 	s_{21}^{(9)}=c_{15}, & 	s_{22}^{(9)}=c_{2}, & 	s_{23}^{(9)}=c_{3}, \\
s_{24}^{(9)}=c_{26}, & 	s_{25}^{(9)}=c_{27}, & 	s_{26}^{(9)}=-c_{24}, & 	s_{27}^{(9)}=-c_{25}, \\	s_{28}^{(9)}=c_{8}, & 	s_{29}^{(9)}=c_{9}, & 	s_{30}^{(9)}=c_{12}, & 	s_{31}^{(9)}=c_{13}, \\
\end{matrix}
\end{array}
$
\end{adjustbox}
\end{center}

\begin{center}
\begin{adjustbox}{width=0.57\textwidth, margin=0 0 0 0}
$
\begin{array}{cccc}
\renewcommand*{\arraystretch}{1.5}
\begin{matrix}
s_{0}^{(10)}=-c_{20}, & 	s_{1}^{(10)}=-c_{21}, & 	s_{2}^{(10)}=-c_{24}, & 	s_{3}^{(10)}=-c_{25}, \\
s_{4}^{(10)}=-c_{8}, & 	s_{5}^{(10)}=-c_{9}, & 	s_{6}^{(10)}=-c_{28}, & 	s_{7}^{(10)}=-c_{29}, \\
s_{8}^{(10)}=c_{4}, & 	s_{9}^{(10)}=c_{5}, & 	s_{10}^{(10)}=-c_{30}, & 	s_{11}^{(10)}=-c_{31}, \\	s_{12}^{(10)}=c_{16}, & 	s_{13}^{(10)}=c_{17}, & 	s_{14}^{(10)}=c_{18}, & 	s_{15}^{(10)}=c_{19}, \\
s_{16}^{(10)}=-c_{12}, & 	s_{17}^{(10)}=-c_{13}, & 	s_{18}^{(10)}=-c_{14}, & 	s_{19}^{(10)}=-c_{15}, \\	s_{20}^{(10)}=c_{0}, & 	s_{21}^{(10)}=c_{1}, & 	s_{22}^{(10)}=-c_{26}, & 	s_{23}^{(10)}=-c_{27}, \\
s_{24}^{(10)}=c_{2}, & 	s_{25}^{(10)}=c_{3}, & 	s_{26}^{(10)}=c_{22}, & 	s_{27}^{(10)}=c_{23}, \\	s_{28}^{(10)}=c_{6}, & 	s_{29}^{(10)}=c_{7}, & 	s_{30}^{(10)}=c_{10}, & 	s_{31}^{(10)}=c_{11}, \\
\end{matrix}
\end{array}
$
\end{adjustbox}
\end{center}

\begin{center}
\begin{adjustbox}{width=0.57\textwidth, margin=0 0 0 0}
$
\begin{array}{cccc}
\renewcommand*{\arraystretch}{1.5}
\begin{matrix}
s_{0}^{(11)}=-c_{23}, & 	s_{1}^{(11)}=-c_{22}, & 	s_{2}^{(11)}=-c_{19}, & 	s_{3}^{(11)}=-c_{18}, \\	s_{4}^{(11)}=c_{11}, & 	s_{5}^{(11)}=c_{10}, & 	s_{6}^{(11)}=-c_{17}, & 	s_{7}^{(11)}=-c_{16}, \\
s_{8}^{(11)}=-c_{31}, & 	s_{9}^{(11)}=-c_{30}, & 	s_{10}^{(11)}=-c_{5}, & 	s_{11}^{(11)}=-c_{4}, \\
s_{12}^{(11)}=-c_{29}, & 	s_{13}^{(11)}=-c_{28}, & 	s_{14}^{(11)}=-c_{25}, & 	s_{15}^{(11)}=-c_{24}, \\
s_{16}^{(11)}=c_{7}, & 	s_{17}^{(11)}=c_{6}, & 	s_{18}^{(11)}=c_{3}, & 	s_{19}^{(11)}=c_{2}, \\	s_{20}^{(11)}=c_{27}, & 	s_{21}^{(11)}=c_{26}, & 	s_{22}^{(11)}=c_{1}, & 	s_{23}^{(11)}=c_{0}, \\
s_{24}^{(11)}=c_{15}, & 	s_{25}^{(11)}=c_{14}, & 	s_{26}^{(11)}=-c_{21}, & 	s_{27}^{(11)}=-c_{20}, \\	s_{28}^{(11)}=c_{13}, & 	s_{29}^{(11)}=c_{12}, & 	s_{30}^{(11)}=c_{9}, & 	s_{31}^{(11)}=c_{8}, \\

\end{matrix}
\end{array}
$
\end{adjustbox}
\end{center}

\begin{center}
\begin{adjustbox}{width=0.57\textwidth, margin=0 0 0 0}
$
\begin{array}{cccc}
\renewcommand*{\arraystretch}{1.5}
\begin{matrix}

s_{0}^{(12)}=-c_{25}, & 	s_{1}^{(12)}=-c_{24}, &  	s_{2}^{(12)}=-c_{21}, & 	s_{3}^{(12)}=-c_{20}, \\	s_{4}^{(12)}=-c_{13}, & 	s_{5}^{(12)}=-c_{12}, & 	s_{6}^{(12)}=-c_{31}, & 	s_{7}^{(12)}=-c_{30}, \\
s_{8}^{(12)}=c_{17}, & 	s_{9}^{(12)}=c_{16}, & 	s_{10}^{(12)}=-c_{29}, & 	s_{11}^{(12)}=-c_{28}, \\	s_{12}^{(12)}=c_{5}, & 	s_{13}^{(12)}=c_{4}, & 	s_{14}^{(12)}=c_{23}, & 	s_{15}^{(12)}=c_{22}, \\
s_{16}^{(12)}=-c_{9}, & 	s_{17}^{(12)}=-c_{8}, & 	s_{18}^{(12)}=-c_{27}, & 	s_{19}^{(12)}=-c_{26}, \\	s_{20}^{(12)}=c_{3}, & 	s_{21}^{(12)}=c_{2}, & 	s_{22}^{(12)}=-c_{15}, & 	s_{23}^{(12)}=-c_{14}, \\
s_{24}^{(12)}=c_{1}, & 	s_{25}^{(12)}=c_{0}, & 	s_{26}^{(12)}=c_{19}, & 	s_{27}^{(12)}=c_{18}, \\	s_{28}^{(12)}=c_{10}, & 	s_{29}^{(12)}=c_{11}, & 	s_{30}^{(12)}=c_{7}, & 	s_{31}^{(12)}=c_{6}, \\
\end{matrix}
\end{array}
$
\end{adjustbox}
\end{center}

\begin{center}
\begin{adjustbox}{width=0.57\textwidth, margin=0 0 0 0}
$
\begin{array}{cccc}
\renewcommand*{\arraystretch}{1.5}
\begin{matrix}
s_{0}^{(13)}=-c_{27}, & 	s_{1}^{(13)}=-c_{26}, & 	s_{2}^{(13)}=-c_{15}, & 	s_{3}^{(13)}=-c_{14}, \\	s_{4}^{(13)}=c_{31}, & 	s_{5}^{(13)}=c_{30}, & 	s_{6}^{(13)}=-c_{13}, & 	s_{7}^{(13)}=-c_{12}, \\
s_{8}^{(13)}=c_{11}, & 	s_{9}^{(13)}=c_{10}, & 	s_{10}^{(13)}=-c_{9}, & 	s_{11}^{(13)}=-c_{8}, \\	s_{12}^{(13)}=c_{7}, & 	s_{13}^{(13)}=c_{6}, & 	s_{14}^{(13)}=c_{3}, & 	s_{15}^{(13)}=c_{2}, \\
s_{16}^{(13)}=c_{29}, & 	s_{17}^{(13)}=c_{28}, & 	s_{18}^{(13)}=c_{25}, & 	s_{19}^{(13)}=c_{24}, \\	s_{20}^{(13)}=-c_{23}, & 	s_{21}^{(13)}=-c_{22}, & 	s_{22}^{(13)}=c_{21}, & 	s_{23}^{(13)}=c_{20}, \\
s_{24}^{(13)}=-c_{19}, & 	s_{25}^{(13)}=-c_{18}, & 	s_{26}^{(13)}=c_{1}, & 	s_{27}^{(13)}=c_{0}, \\	s_{28}^{(13)}=-c_{17}, & 	s_{29}^{(13)}=-c_{16}, & 	s_{30}^{(13)}=-c_{5}, & 	s_{31}^{(13)}=-c_{4}, \\
\end{matrix}
\end{array}
$
\end{adjustbox}
\end{center}

\begin{center}
\begin{adjustbox}{width=0.57\textwidth, margin=0 0 0 0}
$
\begin{array}{cccc}
\renewcommand*{\arraystretch}{1.5}
\begin{matrix}
s_{0}^{(14)}=c_{29}, & 	s_{1}^{(14)}=c_{28}, & 	s_{2}^{(14)}=-c_{31}, & 	s_{3}^{(14)}=-c_{30}, \\	s_{4}^{(14)}=c_{15}, & 	s_{5}^{(14)}=c_{14}, & 	s_{6}^{(14)}=-c_{21}, & 	s_{7}^{(14)}=-c_{20}, \\
s_{8}^{(14)}=-c_{19}, & 	s_{9}^{(14)}=-c_{18}, & 	s_{10}^{(14)}=c_{25}, & 	s_{11}^{(14)}=c_{24}, \\	s_{12}^{(14)}=c_{23}, & 	s_{13}^{(14)}=c_{22}, & 	s_{14}^{(14)}=c_{5}, & 	s_{15}^{(14)}=c_{4}, \\
s_{16}^{(14)}=-c_{27}, & 	s_{17}^{(14)}=-c_{26}, & 	s_{18}^{(14)}=-c_{9}, & 	s_{19}^{(14)}=-c_{8}, \\	s_{20}^{(14)}=-c_{7}, & 	s_{21}^{(14)}=-c_{6}, & 	s_{22}^{(14)}=c_{13}, & 	s_{23}^{(14)}=c_{12}, \\
s_{24}^{(14)}=c_{11}, & 	s_{25}^{(14)}=c_{10}, & 	s_{26}^{(14)}=-c_{17}, & 	s_{27}^{(14)}=-c_{16}, \\	s_{28}^{(14)}=c_{1}, & 	s_{29}^{(14)}=c_{0}, & 	s_{30}^{(14)}=-c_{3}, & 	s_{31}^{(14)}=-c_{2}, \\
\end{matrix}
\end{array}
$
\end{adjustbox}
\end{center}

\begin{center}
\begin{adjustbox}{width=0.57\textwidth, margin=0 10 0 0}
$
\begin{array}{cccc}
\renewcommand*{\arraystretch}{1.5}
\begin{matrix}
s_{0}^{(15)}=-c_{30}, & 	s_{1}^{(15)}=-c_{31}, & 	s_{2}^{(15)}=c_{28}, & 	s_{3}^{(15)}=c_{29}, \\
s_{4}^{(15)}=-c_{26}, & 	s_{5}^{(15)}=-c_{27}, & 	s_{6}^{(15)}=c_{24}, & 	s_{7}^{(15)}=c_{25}, \\
s_{8}^{(15)}=c_{22}, & 	s_{9}^{(15)}=c_{23}, & 	s_{10}^{(15)}=-c_{20}, & 	s_{11}^{(15)}=-c_{21}, \\
s_{12}^{(15)}=-c_{18}, & 	s_{13}^{(15)}=-c_{19}, & 	s_{14}^{(15)}=-c_{16}, & 	s_{15}^{(15)}=-c_{17}, \\
s_{16}^{(15)}=c_{14}, & 	s_{17}^{(15)}=c_{15}, & 	s_{18}^{(15)}=c_{12}, & 	s_{19}^{(15)}=c_{13}, \\	s_{20}^{(15)}=c_{10}, & 	s_{21}^{(15)}=c_{11}, & 	s_{22}^{(15)}=-c_{8}, & 	s_{23}^{(15)}=-c_{9}, \\
s_{24}^{(15)}=-c_{6}, & 	s_{25}^{(15)}=-c_{7}, & 	s_{26}^{(15)}=c_{4}, & 	s_{27}^{(15)}=c_{5}, \\
s_{28}^{(15)}=-c_{2}, & 	s_{29}^{(15)}=-c_{3}, & 	s_{30}^{(15)}=c_{0}, & 	s_{31}^{(15)}=c_{1} \\

\end{matrix}
\end{array}
$
\end{adjustbox}
\end{center}

and

\begin{center}
\begin{adjustbox}{width=0.77\textwidth, margin=0 10 0 10}
$
\begin{array}{cccc}
\renewcommand*{\arraystretch}{1.5}
\begin{matrix}
c_{0}=\frac{1}{2}(b_{0}+b_{1}),   &    c_{1}=\frac{1}{2}(b_{0}-b_{1}), &  c_{2}=\frac{1}{2}(b_{2}+b_{6}),     & c_{3}=\frac{1}{2}(b_{2}-b_{6}), \\
c_{4}=\frac{1}{2}(b_{4}+b_{8}),  &     c_{5}=\frac{1}{2}(b_{4}-b_{8}),	& c_{6}=\frac{1}{2}(b_{3}+b_{7}),     & c_{7}=\frac{1}{2}(b_{3}-b_{7}), \\
c_{8}=\frac{1}{2}(b_{5}+b_{9}),     &  c_{9}=\frac{1}{2}(b_{5}-b_{9}),	& c_{10}=\frac{1}{2}(b_{10}+b_{16}),& c_{11}=\frac{1}{2}(b_{10}-b_{16}), \\
c_{12}=\frac{1}{2}(b_{12}+b_{18}),& c_{13}=\frac{1}{2}(b_{12}-b_{18}),& c_{14}=\frac{1}{2}(b_{14}+b_{20}),& c_{15}=\frac{1}{2}(b_{14}-b_{20}), \\
c_{16}=\frac{1}{2}(b_{11}+b_{17}),& c_{17}=\frac{1}{2}(b_{11}-b_{17}),& c_{18}=\frac{1}{2}(b_{13}+b_{19}),& c_{19}=\frac{1}{2}(b_{13}-b_{19}), \\
c_{20}=\frac{1}{2}(b_{15}+b_{21}), &c_{21}=\frac{1}{2}(b_{15}-b_{21}), &c_{22}=\frac{1}{2}(b_{22}+b_{26}),& c_{23}=\frac{1}{2}(b_{22}-b_{26}), \\
c_{24}=\frac{1}{2}(b_{24}+b_{28}),& c_{25}=\frac{1}{2}(b_{24}-b_{28}), &c_{26}=\frac{1}{2}(b_{23}+b_{27}),& c_{27}=\frac{1}{2}(b_{23}-b_{27}), \\
c_{28}=\frac{1}{2}(b_{25}+b_{29}),& c_{29}=\frac{1}{2}(b_{25}-b_{29}), &c_{30}=\frac{1}{2}(b_{30}+b_{31}),& c_{31}=\frac{1}{2}(b_{30}-b_{31}),
\end{matrix}
\end{array}
$
\end{adjustbox}
\end{center}

Using the above matrices and relations the computational procedure for calculating Kaluza numbers product can be written as follows:

\begin{equation}
\mathbf{Y}_{32\times1}=\mathbf{P}_{32}\mathbf{W}_{32}\mathbf{A}_{32\times512}\mathbf{D}_{512}\mathbf{P}_{512\times32}\mathbf{W}_{32}\mathbf{P}_{32}\mathbf{X}_{32\times1}
\label{6}
\end{equation}

\begin{adjustbox}{margin = 0 0 0 8}
\end{adjustbox}

It is easy to see that the diagonal matrix $\mathbf{D}_{512}$ contains only 32 elements differing by its
value. The remaining 480 elements coincide with these thirty two elements up to a sign. Let us create a column vector $\mathbf{C}_{32\times1}=[ c_0,c_1,\dots,c_{31}]^{\mathrm{T}}$, consisting of the all 32 elements of matrix $\mathbf{D}_{512}$ which possess different values. Then it is easy to see that the elements $ \{ c_k \} $, $k=0,1,\dots,31$ can be calculated using the following vector–matrix procedure:

\[
\mathbf{C}_{32\times1}=\frac{1}{2}(\mathbf{I}_{16}\otimes\mathbf{H}_{2})\mathbf{B}_{32\times1},\mathbf{B}_{32\times1}=[b_0,b_1,\dots,b_{31}]^{\mathrm{T}}.
\]

%\begin{adjustbox}{margin = 0 0 0 8}
%\end{adjustbox}

\begin{figure}[b!]
 \centering
  \includegraphics[width=0.8\textwidth]{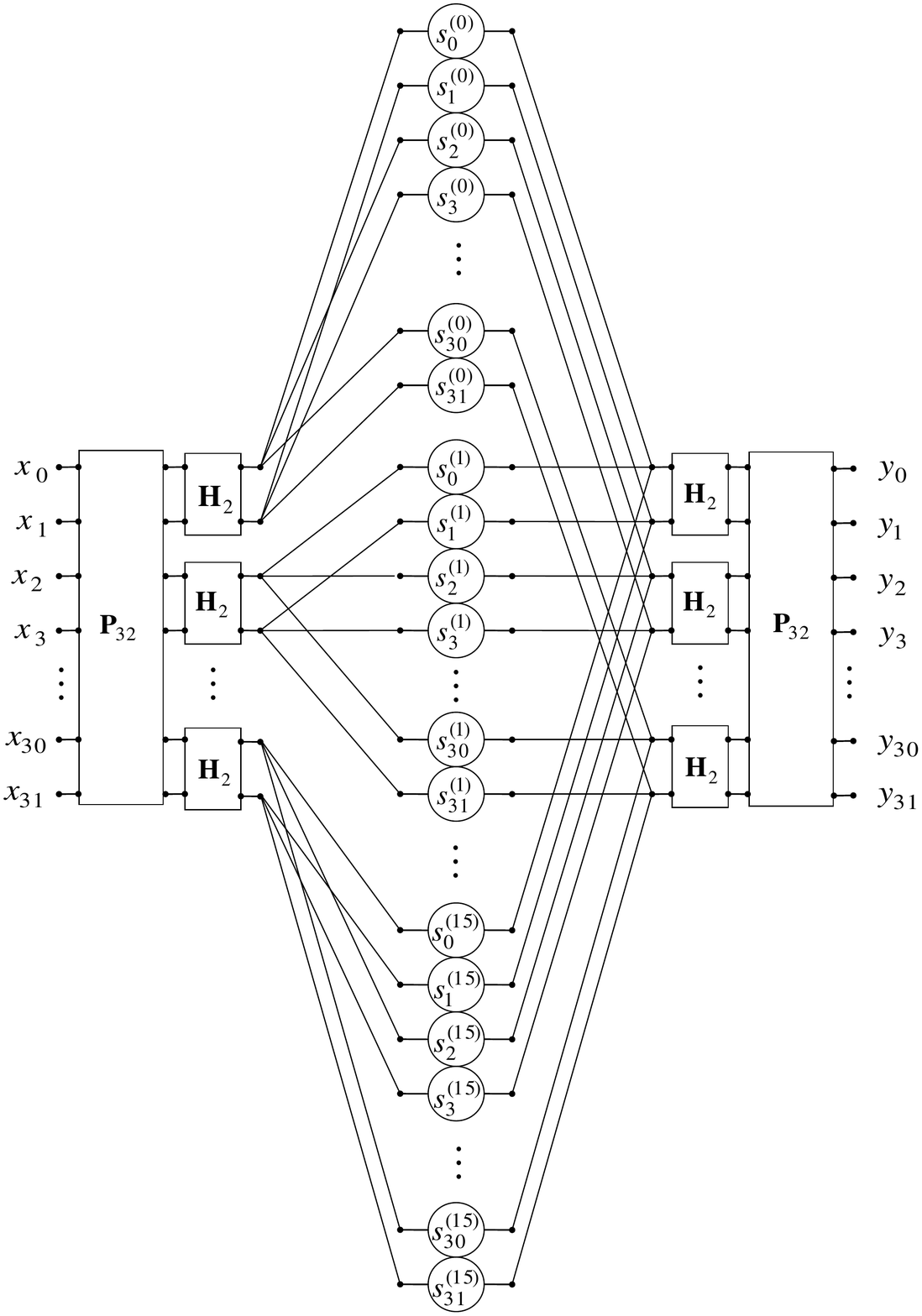}
  \caption{Data flow diagram for rationalized Kaluza numbers multiplication algorithm in accordance
with the procedure (3.3)}
  \label{fig2}
\end{figure}
\FloatBarrier

Fig. 2 shows a data flow diagram representation of the rationalized algorithm for computation
of Kaluza numbers product and Fig. 3 shows a data flow diagram of the process
for calculating the vector $\mathbf{C}_{32\times1}$ elements. In Fig. 2, the points where lines converge denote
summation. As follows from Fig. 3, calculation of elements of diagonal matrix $\mathbf{D}_{512}$ requires
performing only trivial multiplications by the power of two. Such operations may be implemented using convention arithmetic shift operations, which have simple realization and hence may be neglected during computational complexity estimation \cite{24}.

\begin{figure}[hb]
 \centering
  \includegraphics[width=0.25\textwidth]{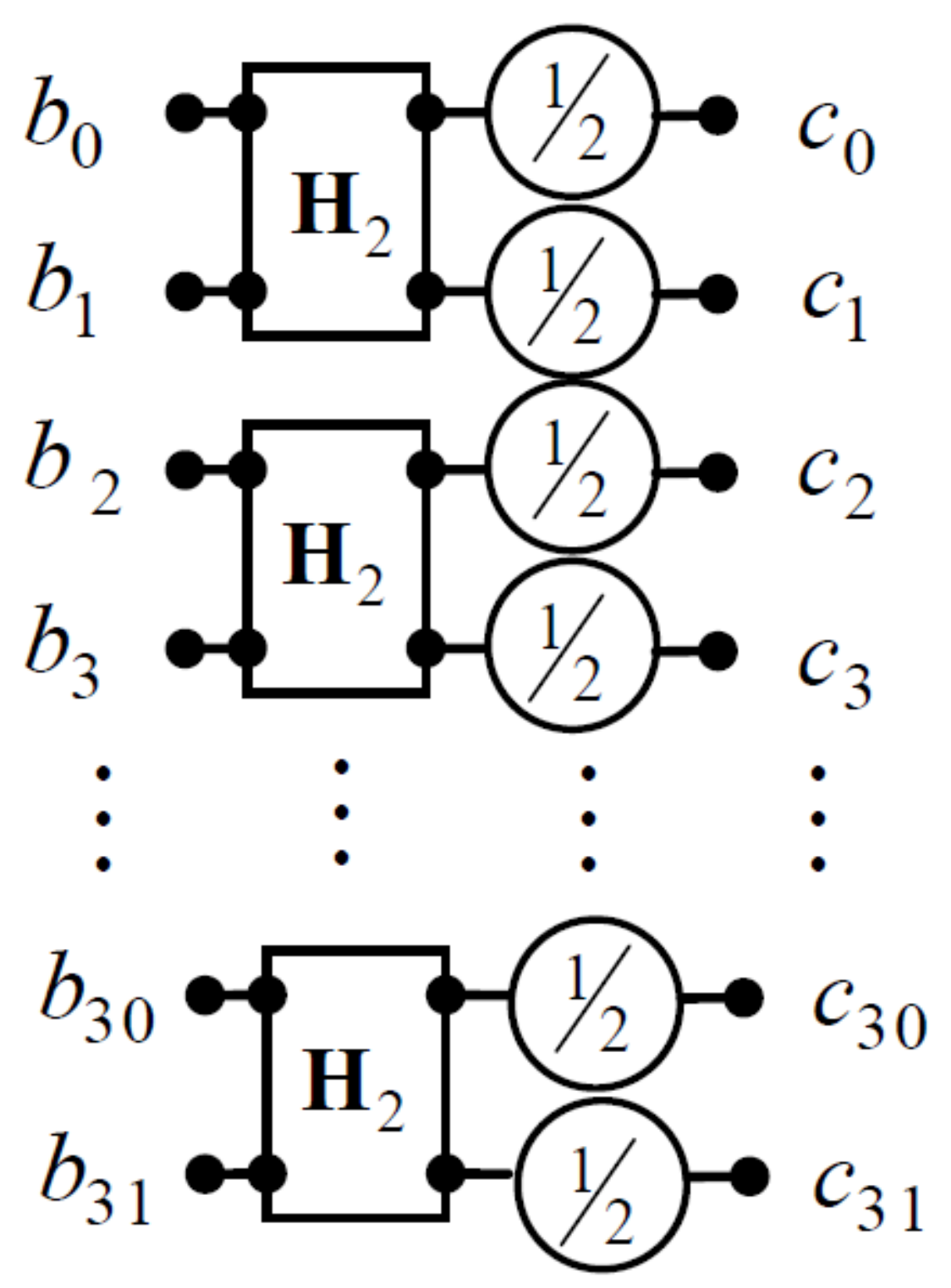}
  \caption{The data flow diagram describing the process of calculating elements of the vector $\mathbf{C}_{32\times1}$ }
  \label{fig3}
\end{figure}
\FloatBarrier

\section{Evaluation of computational complexity}

We calculate how many real multiplications (excluding multiplications by power of two)
and real additions are required for realization of the proposed algorithm, and compare it
with the number of operations required for a direct evaluation of matrix–vector product in
Eq. (2.3). As already mentioned the number of real multiplications required using the proposed
algorithm is 512. Thus using the proposed algorithm the number of real multiplications
to calculate Kaluza number product is halved. The number of real additions required
using our algorithm is 576. Therefore, the total number of arithmetic operations for proposed
algorithm is approximately 46\% less than that of the direct evaluation.

\section{Conclusions}

In this paper, we have presented an original algorithm allowing to multiply Kaluza numbers with reduced both multiplicative and additive complexity. Furthermore, the total number of operations in our algorithm is almost two times less than the total number of operations in the compared algorithm. Therefore, the proposed algorithm is better than the naive algorithm, both in terms of hardware implementation and in terms of its software implementation on a conventional computer too.
\\
\\
\\
\\

% ------------------------------------------------------------------------
\end{document}